\documentclass[11pt]{article}%
\usepackage{amsfonts}
\usepackage{xcolor}
\colorlet{linkequation}{green}
\usepackage{amsmath}%
\usepackage{amssymb}%
\usepackage{graphicx}
\usepackage{float}
\usepackage{longtable}
\usepackage[utf8]{inputenc}
\usepackage{hyperref}
\usepackage{natbib} 
\usepackage{mathdots}
\usepackage{upgreek}
\usepackage{algorithm,algcompatible}
\usepackage{amsmath}

\usepackage{enumerate}

\newcommand{\fref}[1]{Fig.~\ref{#1}}

\newcommand{\tref}[1]{Table~\ref{#1}}
\newcommand{\sref}[1]{Section~\ref{#1}}

\newcommand*{\cref}[1]{%
  \begingroup
    \hypersetup{
      linkcolor=linkequation,
      linkbordercolor=linkequation,
    }%
    \ref{#1}%
  \endgroup
}

\providecommand{\U}[1]{\protect\rule{.1in}{.1in}}
\newtheorem{theorem}{Theorem}

\newtheorem{lemma}{Lemma}

\newenvironment{proof}[1][Proof]{\textbf{#1.} }{\ \rule{0.5em}{0.5em}}
\begin{document}

\title{Quantum Imaging of High-Dimensional Hilbert Spaces with Radon Transform}
\author{Laszlo Gyongyosi$^{1,2,3,}$\thanks{E-mail: \href{mailto:l.gyongyosi@soton.ac.uk}{l.gyongyosi@soton.ac.uk}}\\
$^{1}$School of Electronics and Computer Science\\University of Southampton\\Southampton SO17 1BJ, UK\\
$^{2}$Department of Networked Systems and Services\\Budapest University of Technology and Economics\\Budapest, H-1117 Hungary\\
$^{3}$MTA-BME Information Systems Research Group\\Hungarian Academy of Sciences\\Budapest, H-1051 Hungary}
\date{}

\maketitle
\begin{abstract}
High-dimensional Hilbert spaces possess large information encoding and transmission capabilities. Characterizing exactly the real potential of high-dimensional entangled systems is a cornerstone of tomography and quantum imaging. The accuracy of the measurement apparatus and devices used in quantum imaging is physically limited, which allows no further improvements to be made. To extend the possibilities, we introduce a post-processing method for quantum imaging that is based on the Radon transform and the projection-slice theorem. The proposed solution leads to an enhanced precision and a deeper parameterization of the information conveying capabilities of high-dimensional Hilbert spaces. We demonstrate the method for the analysis of high-dimensional position-momentum photonic entanglement. We show that the entropic separability bound in terms of standard deviations is violated considerably more strongly in comparison to the standard setting and current data processing. The results indicate that the possibilities of the quantum imaging of high-dimensional Hilbert spaces can be extended by applying appropriate calculations in the post-processing phase.
\end{abstract}

\section{Introduction}
\label{sec1}
The field of quantum imaging has been proposed to reveal and exploit the deeply involved, currently open and still uncharacterized hidden potentials of quantum mechanics. Quantum imaging has already been applied successfully in the fields of quantum optics, ghost imaging, quantum lithography and quantum sensing [\cref{ref1}-\cref{ref28}]. One of the most interesting subfields of quantum imaging is related to the study of high-dimensional entangled spaces [\cref{ref1}, \cref{ref7}]. High-dimensional Hilbert spaces represent a useful resource for quantum computation, quantum communication protocols and quantum cryptography. A high-dimensional entangled system is equipped with several important features and offers numerous additional benefits, but for communication purposes one of the most important properties is the large data encoding and transmission capability. In particular, the exact characterization of the information-conveying property of a high-dimensional Hilbert space is a cornerstone of quantum imaging and tomography. A photonic entangled system can convey several bits in a single photon state. One of the most plausible members of this set is the high-dimensional \textit{position-momentum entanglement}, since the photonic position and momentum degree of freedom can be efficiently manipulated within the current technological framework. Another tangible example of position-momentum coding is continuous-variable quantum communications, where the information is encoded into the position and momentum quadratures of the coherent states. The transmission capability of these kinds of high-dimensional Hilbert spaces can be quantified by the photon coincidence detections, whose measurement data finally ``draw an image'' from the exploitable encoding possibilities of the analyzed space. On the other hand, the accuracy of quantum imaging is limited by several factors, most importantly by imperfections of the measurement process and the fundamental laws of quantum mechanics. Since these boundaries and limitations cannot be neglected, an appropriate solution for the enhancement of the current solutions could be only the application of clever data processing steps and numerical calculations on the measured \textit{raw data} in the so-called \textit{post-processing} phase. The post-processing phase operates on the raw data that is resulted from the coincidence detections in the physical layer, and requires no further quantum-level interactions, i.e., the problem of enhancing can be converted and reformulated from the physical layer into the logical layer. All further steps that are related to any boosting operations will be made in this layer, which is a particularly convenient approach, since we get a ``free hand'' to maximize the extractable valuable information from the raw data by any intelligent computational steps.

The information transmission capability of a high-dimensional entangled quantum system can be rephrased in the framework and well-known tools of quantum Shannon theory [\cref{ref29}, \cref{ref33}-\cref{ref35}]. High-dimensional position-momentum entanglement [\cref{ref1}, \cref{ref7}] can also be discussed by appropriate correlation measure functions of this field, such as the mutual information that quantifies exactly the classical correlation of two quantum systems -- such as between the subsystems of a high-dimensional entangled biphotonic system. Taking into consideration the joint coincidence detections in the measurement process, the mutual information function is an appropriate measure to study and quantify precisely the information transmission capabilities of high-dimensional Hilbert spaces. In particular, the mutual information function in the level of the logical layer results from joint photon coincidence detection events in the level of the physical layer. Hence there is a strict connection between the physical layer and the mutual information function that specifically derives from these measurement data. The accuracy of the measurement apparatus is critical, and unfortunately it is also limited by the laws of quantum mechanics. An appropriate answer could be to integrate some ``intelligence'' into the post-processing phase, which can be applied freely on the raw data to extract as much valuable information as possible. Since the physical limitations of the measurement process cannot be traced out from the picture, only one path remains to enhance the performance and quality: to find an appropriate post-processing in the logical layer. All improvement has to be investigated and integrated into this layer. 

In the process of quantum imaging of high-dimensional photonic entanglement, the information transmission capability is characterized by coincidence detections. The statistic of the joint detection events builds up the mutual information function, which finally leads to an adequate description of the information transmission capability of the Hilbert space. The measurement devices (practically controllable pixel mirrors) are equipped with a given measurement dimension (referred as measurement space or resolution). Since in current quantum imaging and tomography several imperfections are added into the process, the detection is \textit{not optimal}. Having arrived at this point, according to these argumentations our answer has to be clear: post-processing. Numerical post-processing techniques have already demonstrated their capability in several different areas related to quantum computations and communications, and have been found to be a useful tool in enhancing and amplifying the performance of physical layer processes. A carefully constructed post-processing consists of several algorithmical steps, and basically it is performed by purely the logical layer, i.e., in an abstract layer independently from the physical layer. It also means that no further physical interaction is needed to improve the performance of the analyzed system. As we have found, it is also possible to boost the capabilities of quantum imaging and to achieve more accurate and precise results from the collected coincidence measurements by applying an appropriate post-processing technique in the data processing phase. The \textit{effective entropic channel} \textit{quantity} [\cref{ref1}, \cref{ref7}] takes into consideration both the entangled photonic system and the measurement apparatus. It is a suitable measure to quantify accurately the transmission capacity of the high-dimensional photonic Hilbert space in \textit{bits of information per photon}. The entropic channel measure is analogous to the Shannon capacity formula and is characterized by several joint photon detection events [\cref{ref1}]. Briefly, our model uses the mutual information function, and our approach also lies on the use of this essential quantity. 

The Radon transform is a useful tool in medical imaging and particularly in the processes of medical tomography. This transform consists of the integral transform of several pieces of an unknown function (e.g., a physical object), from which the unknown density can be recovered by the inverse Fourier transform. A well-known medical application of Radon transform is X-raying, where several parallel lines (rays) each from a different angle convey information about an unknown internal density function, and each ray captures and characterizes a different piece of the unknown target. The aim of Radon transform in these traditional applications is to collect together these information slices, and then to apply an appropriate inverse transformation that is able to recover the unknown internal function from the gathered slices. In our quantum imaging scenario we explicitly do the same thing to reach several advantageous features. However, instead of physically emitted rays and spatial rotations (such as is the case in X-raying), our model is interpreted by ``abstracted'' lines in the high-dimensional Hilbert space, whose ``lines'' are defined by the coincidence measurements and convey information about the position and momentum components of the analyzed high-dimensional quantum system. Similarly, the rotation does not mean a physical rotation in the spatial space, but a unitary transformation in the phase, as will be revealed in detail in \sref{sec3}.

In this work, we introduce a Radon transform-based post-processing for quantum imaging and quantum tomography, which uses the raw data of the coincidence measurements to enhance the accuracy of the study of information transmission capabilities of high-dimensional position-momentum entangled quantum states. The proposed post-processing phase provides several benefits for us to get a sharper and considerably deeper picture from the internal life of high-dimensional Hilbert spaces, without the necessity of any further quantum-level interactions in the physical layer. 

This paper is organized as follows. In \sref{sec2}, the preliminaries are summarized. \sref{sec3} discusses the proposed scheme, while \sref{sec4} reveals the data processing steps. Finally, \sref{sec5} concludes the paper. Supplementary material is included in the Appendix.
 
\section{Preliminaries}
\label{sec2}
 The information transmission capability of the high-dimensional position-momentum entangled photonic state will be quantified by photon coincidence detections, which lead to the direct application of the mutual information function. 

 The mutual information between discrete variables $A$ and $B$ is denoted by $I\left(A\mathrm{:}B\right)$, and given by $I\left(A\mathrm{:}B\right)\mathrm{=}H\left(A\right)\mathrm{+}H\left(B\right)\mathrm{-}H\left(AB\right)$, where $H\left(\mathrm{\cdot }\right)$ is the Shannon entropy, $H\left(A\right)\mathrm{=}\sum_{x\mathrm{\in }A}{p\left(x\right)}\mathrm{log}p\left(x\right)$, while the joint entropy is as $H\left(AB\right)\mathrm{=}\sum_{ \begin{array}{l}
x\mathrm{\in }A, y\mathrm{\in }B \end{array}
}{p\left(x,y\right)}\mathrm{log}p\left(x,y\right)$.

 A position-momentum entangled photonic bipartite state $\left|\left.{\psi }_{AB}\right\rangle \right.$ can be characterized by the entangled biphoton wave function in the position and momentum basis, respectively, as follows. Introducing the notations $x_A$ and $x_B$ for the position basis, the biphoton wave function $f\left(x_A,x_B\right)$ is expressed as [\cref{ref1}]
\begin{equation} \label{1)} 
f\left(x_A,x_B\right)\mathrm{=}Ne^{\frac{\mathrm{-}{\left(x_A\mathrm{-}x_B\right)}^{\mathrm{2}}}{\mathrm{4}w^{\mathrm{2}}_{\mathrm{1}}}}e^{\frac{\mathrm{-}{\left(x_A\mathrm{+}x_B\right)}^{\mathrm{2}}}{\mathrm{16}w^{\mathrm{2}}_{\mathrm{2}}}}, 
\end{equation} 
where 
\begin{equation} \label{2)} 
N\mathrm{=}\frac{\mathrm{1}}{\mathrm{2}\pi w_{\mathrm{1}}w_{\mathrm{2}}},                         
\end{equation} 
while $\mathrm{2}w_{\mathrm{1}}$ is the Gaussian width in the $x_{\mathrm{1}}\mathrm{-}x_{\mathrm{2}}$ direction, and $w_{\mathrm{2}}$ is the Gaussian width in the $x_{\mathrm{1}}\mathrm{+}x_{\mathrm{2}}$ direction [\cref{ref1}, \cref{ref7}]. In the momentum basis, the biphoton wave function $f\left(p_A,p_B\right)$ is evaluated as 
\begin{equation} \label{3)} 
f\left(p_A,p_B\right)\mathrm{=}{\left(\mathrm{4}w_{\mathrm{1}}w_{\mathrm{2}}\right)}^{\mathrm{2}}Ne^{\mathrm{-}w^{\mathrm{2}}_{\mathrm{1}}{\left(p_A\mathrm{-}p_B\right)}^{\mathrm{2}}}e^{\mathrm{-}\mathrm{4}w^{\mathrm{2}}_{\mathrm{2}}{\left(p_A\mathrm{+}p_B\right)}^{\mathrm{2}}}. 
\end{equation} 
From $w_{\mathrm{1}}$ and $w_{\mathrm{2}}$, the measured single photon width ${\sigma }_s$ is expressed as
\begin{equation} \label{ZEqnNum574414} 
{\sigma }^{\mathrm{2}}_s\mathrm{=}w^{\mathrm{2}}_{\mathrm{2}}\mathrm{+}{\left(\frac{w_{\mathrm{1}}}{\mathrm{2}}\right)}^{\mathrm{2}},                 
\end{equation} 
while the conditional width ${\sigma }_C$ is as follows:
\begin{equation} \label{ZEqnNum516221} 
{\sigma }^{\mathrm{2}}_C\mathrm{=}\frac{\mathrm{4}w^{\mathrm{2}}_{\mathrm{1}}w^{\mathrm{2}}_{\mathrm{2}}}{\mathrm{4}w^{\mathrm{2}}_{\mathrm{2}}\mathrm{+}w^{\mathrm{2}}_{\mathrm{1}}}.       
\end{equation} 
Assuming that $w_{\mathrm{1}}\mathrm{\ll }w_{\mathrm{2}}$ holds, these relations are simplified to
\begin{equation} \label{6)} 
{\sigma }_s\mathrm{=}w_{\mathrm{2}} 
\end{equation} 
and
\begin{equation} \label{7)} 
{\sigma }_C\mathrm{=}w_{\mathrm{1}}.              
\end{equation} 
Further details about the characterization of these functions can be found in [\cref{ref1}, \cref{ref7}].
 
\section{Quantum Imaging with Radon Transform}
\label{sec3}
 Radon transform is a well-known and applied technique in medical tomography to discover an unknown two-dimensional internal density function $\mu \left(x,y\right)$, where $x$ and $y$ are variable parameters. (An illustrative example of the application of the Radon transform is in X-raying, where a two-dimensional picture is constructed from the unknown density function.) 

 In the traditional interpretation (i.e., for non-quantum imaging purposes) of Radon transform, the task is to recover $\mu \left(x,y\right)$ from the knowledge of the measurement (such as the light intensity) results. The information about the unknown internal function is divided into several parallel lines, each conveying partial information or slice of information about the unknown function. Radon transform integrates these slices together to extract and recover the full information about the unknown function. 
In practice, these Radon transform steps are as follows. All information that could be cumulated from an unknown function $\mu \left(x,y\right)$ across a single path can be described by an appropriate integral operation. Taking a line $L$ through the unknown density function, the line integral of $\mu $ along $L$ can be expressed as:
\begin{equation} \label{8)} 
\int_L{\mu \left(x,y\right)}da,               
\end{equation} 
where $a$ is the arc length parameter [\cref{ref30}-\cref{ref32}]. Since, by the nature of the problem, it is not possible to fully recover $\mu \left(x,y\right)$ from a single line \textit{L}, the tomography process has to take into account several other paths each from a different angle $\phi \mathrm{,0}\mathrm{\le }\phi \mathrm{<}\pi $. Each path catches and characterizes a different property of the unknown density function. In particular, one can obtain several different line integrals through the unknown density to build up a detailed picture, hence the main task is to determine the unknown density function $\mu \left(x,y\right)$ from the measured line integrals and the variable density function values. A given path $\Omega$ can convey only partial information about the internal function, and can be modeled as
\begin{equation} \label{ZEqnNum142582} 
\Omega\left(\xi \right)\mathrm{=}\int^{\xi }_{{\xi }_0}{\mu \left(x,y\right)}dxdy,            
\end{equation} 
where ${\xi }_0,\xi $ are points of the line $L$. The unknown function $\mu \left(x,y\right)$ can be computed from the derivate $\Omega\mathrm{''}$, however it requires the full knowledge of \eqref{ZEqnNum142582}, which is not a reasonable assumption in any practical scenario. Hence, the appropriate calculation requires the use of several different rotation angles $\phi $ (i.e., a sensor rotates about a center, a plausible example for this is X-raying.). Fortunately, in our setting this kind of spatial restriction can be removed and the formula of \eqref{ZEqnNum142582} can be directly applied, however some further steps are still needed to apply it in the quantum imaging.

 At this point, we have to turn our attention from the traditional interpretation to the quantum imaging of high-dimensional Hilbert spaces, specifically the position-momentum entanglement. Fortunately, a well-characterized connection exists between them. In our quantum imaging scenario, the unknown two-dimensional function identify an\textit{ mutual information} \textit{slice} as
\begin{equation} \label{ZEqnNum264887} 
\mu \left(x,p\right),                            
\end{equation} 
where \textit{x} and \textit{p} are the position and momentum components. (For the exact derivation of the mutual information function in a Radon transform of a high-dimensional entangled system, see \sref{sec31}.) 

\begin{figure*}[h!]
 \begin{center}
 	 \includegraphics[angle = 0,width=1\linewidth]{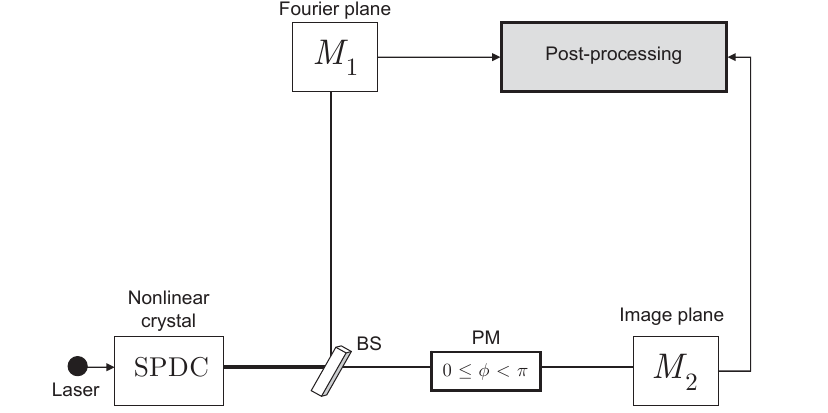}

\caption{The measurement setup for the Radon transform-based quantum imaging. The output of the nonlinear crystal is fed into a beamsplitter (BS). The outputs of the BS are measured in the Fourier plane and in the image plane. The image plane path also contains a phase modulator (PM) for the unitary phase rotation. The measurements are taken for $\mathrm{0}\mathrm{\le }\phi \mathrm{<}\pi $. In the post-processing phase only the coincident photon detections are taken into account to derive the mutual information.}
 \label{fig1}
\end{center}
\end{figure*}

 In terms of the measured raw data the encoded mutual information quantities as follows. A given \textit{i}-th coincidence detection measurement $M_{\phi ,i}$ at a given \textit{phase delay} $\phi \mathrm{,0}\mathrm{\le }\phi \mathrm{<}\pi $ (see \fref{fig1}) defines an \textit{encoded information slice}
\begin{equation} \label{ZEqnNum349368} 
\mathcal{E}\left(\mu \left(x,p\right)\right)\mathrm{=}\int_{M_{\phi ,i}}{\mu \left(x,p\right)}dxdp,                                      
\end{equation} 
which conveys a piece from the mutual information that can be extracted from the position and momentum components, respectively. 

 Putting \textit{n} encoded slices of \eqref{ZEqnNum349368} together and freezing the phase delay into $\phi $ leads to the \textit{encoded partial mutual information} 
\begin{equation} \label{ZEqnNum787631} 
\mathcal{E}\left(I_{M_{\phi }}\left(A\mathrm{:}B\right)\right)=\mathcal{E}\left({\mu }_{\phi }\left(x,p\right)\right)\mathrm{=}\int_{M_{\phi \mathrm{,1}}}{\mathrm{\dots }}\int_{M_{\phi ,n}}{\mu \left(x,p\right)}dxdp,              
\end{equation} 
which information is present in the form of the coincidence detections $M_{\phi }$ at a given value of $\phi $. 

 The \textit{encoded} \textit{full mutual information}, $\mathcal{E}\left(I_{\mathcal{R}}\left(A\mathrm{:}B\right)\right)$ that is contained in the raw data is evaluated as
\begin{equation} \label{ZEqnNum858033} 
 \begin{array}{l}
\begin{split}
\mathcal{E}\left(I_{\mathcal{R}}\left(A\mathrm{:}B\right)\right)&\mathrm{=}\int_{\phi }{\mathcal{E}\left(I_{M_{\phi }}\left(A\mathrm{:}B\right)\right)}d\phi  \\ 
&\mathrm{=}\int_{\phi }{\mathcal{E}\left({\mu }_{\phi }\left(x,p\right)\right)}d\phi  \\ 
&\mathrm{=}\int_{\phi }{\int_{M_{\phi \mathrm{,1}}}{\mathrm{\dots }}\int_{M_{\phi ,n}}{\mu \left(x,p\right)}dxdp}d\phi,
\end{split} 
\end{array}
\end{equation} 
where $\mathrm{0}\mathrm{\le }\phi \mathrm{<}\pi $. The task is to recover the full mutual information function $I_{\mathcal{R}}\left(A\mathrm{:}B\right)$ from the knowledge of the raw data values $\mathcal{E}\left(I_{\mathcal{R}}\left(A\mathrm{:}B\right)\right)$. As one can readily see, with no phase delay $\phi $ (i.e., $\phi \mathrm{=0}$), a single coincidence measurement precisely leads to the mutual information $I_0\left(A\mathrm{:}B\right)$ in \eqref{ZEqnNum787631}, which is exactly the case in a standard setting. 

 The Radon transform-based quantum imaging builds up the mutual information function $I_{\mathcal{R}}\left(A\mathrm{:}B\right)$ from several different fractions where each fraction, in fact, conveys a partial mutual information function $I_{M_{\phi }}\left(A\mathrm{:}B\right)$. 

 The $\mathcal{R}$ Radon transform of the unknown slice $\mu \left(\rho ,\phi \right)$ can be expressed as 
\begin{equation} \label{ZEqnNum312017} 
 \begin{array}{l}
\begin{split}
\mathcal{R}\left(\mu \left(\rho ,\phi \right)\right)&=\mathcal{E}\left(\mu \left(x,p\right)\right) \\ 
&\mathrm{=\ }\int_{M_{\phi ,i}}\mu \left(x,p\right)dxdp, 
\end{split}
\end{array}                            
\end{equation} 
which identifies an information slice (see \eqref{ZEqnNum349368}), where $\rho $ is defined as
\begin{equation} \label{ZEqnNum520700} 
\rho \mathrm{=}\left(x,p\right)\mathrm{\cdot }\left(\mathrm{cos}\phi \mathrm{,sin}\phi \right)\mathrm{=}x\mathrm{cos}\phi \mathrm{+}p\mathrm{sin}\phi .            
\end{equation} 
For the \textit{i}-th abstracted line, for $\mathrm{0}\mathrm{\le }\phi \mathrm{<}\pi $, 
\begin{equation} \label{ZEqnNum274893} 
\rho _{i} =\left(x_{i} ,p_{i} \right)\cdot \left(\cos \phi ,\sin \phi \right)=x_{i} \cos \phi +p_{i} \sin \phi,                 
\end{equation}
where $x_i,p_i$ are the position and momentum components that identify a slice of the partial mutual information function $I_{M_{\phi }}\left(A\mathrm{:}B\right)$ at a given $\phi $. At a fixed $\phi $, the collection of \textit{n} parameters of \eqref{ZEqnNum274893} each belong to a given slice $\mu \left(x_i,p_i\right)$ is referred by ${\rho }_{\phi }$.

 The partial and full mutual information functions (conveyed in the raw data) are evaluated by the Radon transform of unknown functions ${\mu }_{\phi }\left({\rho }_{\phi },\phi \right)$ and $\int_{\phi }{{\mu }_{\phi }\left({\rho }_{\phi },\phi \right)d\phi }$ as follows:
\begin{equation} \label{ZEqnNum813599} 
 \begin{array}{l}
\begin{split}
\mathcal{R}\left({\mu }_{\phi }\left({\rho }_{\phi },\phi \right)\right) 
&=\mathcal{E}\left({\mu }_{\phi }\left(x,p\right)\right) \\ 
&\mathrm{=\ }\int_{M_{\phi \mathrm{,1}}}{\mathrm{\dots }}\int_{M_{\phi ,n}}{\mu \left(x,p\right)}dxdp 
\end{split}
\end{array}
\end{equation} 
and
\begin{equation} \label{ZEqnNum200425} 
 \begin{array}{l}
\begin{split}
\mathcal{R}\left(\int_{\phi }{{\mu }_{\phi }\left({\rho }_{\phi },\phi \right)d\phi }\right)  
&=\mathcal{E}\left(\int_{\phi }{{\mu }_{\phi }\left(x,p\right)d\phi }\right) \\ 
&\mathrm{=\ }\int_{\phi }{\int_{M_{\phi \mathrm{,1}}}{\mathrm{\dots }}\int_{M_{\phi ,n}}{\mu \left(x,p\right)}dxdpd\phi }. 
\end{split}
\end{array}
\end{equation} 
As one can readily conclude, the results of coincidence detections in \eqref{ZEqnNum349368}, \eqref{ZEqnNum787631} and \eqref{ZEqnNum858033} can be reformulated as a Radon transform shown in \eqref{ZEqnNum312017}, \eqref{ZEqnNum813599} and \eqref{ZEqnNum200425}. For \eqref{ZEqnNum520700}, a function $\delta $ can be introduced along with the Cartesian equation $\rho \mathrm{-}x\mathrm{cos}\phi \mathrm{+}p\mathrm{sin}\phi \mathrm{=0}$. This function is referred to as
\begin{equation} \label{ZEqnNum286882} 
\delta \left(\rho \mathrm{-}x\mathrm{cos}\phi \mathrm{-}p\mathrm{sin}\phi \right),                    
\end{equation} 
and is called the line impulse in the standard interpretation [\cref{ref31}, \cref{ref32}]. 

 The schematic view of the measurement setup for the Radon transform-based high-dimensional quantum imaging is summarized in \fref{fig1}. The source is assumed to be a collimated laser beam that has undergone a spontaneous parametric down-conversion (SPDC) at a nonlinear crystal. The outputs of the BS are sent to micro-mirror devices, at the Fourier plane and the image plane. The unitary phase rotation of $\phi $ is implemented by a phase modulator (PM) in the image plane path. Other supplementary devices (focusing lens, quarter wave plates, polarizing beam splitters) of the experimental setting are not depicted in the figure and are not part of our discussion, these can be found in the literature [\cref{ref1}-\cref{ref7}]. The detectors are characterized by their dimension (measurement space), $d$, and the capacity of the quantum system is measured in ${\mathrm{bits}}/{\mathrm{photon}}$, which is quantified by the joint detection events in the coincidence measurement. (\textit{Note}: A general setup [\cref{ref1}, \cref{ref7}] contains no PM, i.e., $\phi \mathrm{=0}$, this setup is referred to as the \textit{standard model} throughout.) 

\begin{theorem} 
(Projection-slice theorem for quantum imaging). The $\mathcal{R}\left({\mu }_{\phi }\right)$\textit{ Radon transform of }${\mu }_{\phi }\left(\rho ,\phi \right)$\textit{ leads to the two-dimensional Fourier transform }$F$\textit{ of }${\mu }_{\phi }$\textit{. Taking the two-dimensional inverse Fourier transform of }$\mathcal{R}\left({\mu }_{\phi }\right)$\textit{ results in the partial mutual information function }$F^{-1}\left(\mathcal{R}\left({\mu }_{\phi }\right)\right)={\mu }_{\phi }=I_{M_{\phi }}\left(A:B\right)$\textit{. Evaluating the two-dimensional inverse Fourier transform for the full domain of }$\phi $\textit{ leads to the mutual information as }$F^{-1}_{0\le \phi <\pi }\left(\mathcal{R}\left({\mu }_{0\le \phi <\pi }\right)\right)=I_{\mathcal{R}}\left(A:B\right)$\textit{.}
\end{theorem}
\begin{proof} 
 Using \eqref{ZEqnNum286882}, the Radon transform of the partial mutual information function in \eqref{ZEqnNum813599} can be evaluated as
\begin{equation} \label{20)} 
\mathcal{R}\left({\mu }_{\phi }\left({\rho }_{\phi },\phi \right)\right)\mathrm{=}\int^{\mathrm{\infty }}_{\mathrm{-}\mathrm{\infty }}{\int^{\mathrm{\infty }}_{\mathrm{-}\mathrm{\infty }}{\mu \left(x,p\right)}}\delta \left({\rho }_{\phi }\mathrm{-}x\mathrm{cos}\phi \mathrm{-}p\mathrm{sin}\phi \right)dxdp\mathrm{\ .} 
\end{equation} 
First we show that the Fourier transform of $\mathcal{R}\left({\mu }_{\phi }\right)$ with respect to variable ${\rho }_{\phi }$ at a $\phi $ fixed value, denoted by $F_{{\rho }_{\phi }}\left(\mathcal{R}\left({\mu }_{\phi }\right)\right)$, is in fact equal to the two-dimensional Fourier transform of the partial mutual information function ${\mu }_{\phi }\left(x,p\right)$\textit{.} 

 To evaluate it, we introduce the dual variable of ${\rho }_{\phi }$, referred as the imaginary frequency variable $r$. Then,
\begin{equation} \label{ZEqnNum935273} 
 \begin{array}{l}
\begin{split}
&F_{{\rho }_{\phi }}\left(\mathcal{R}\left({\mu }_{\phi }\right)\left(r,\phi \right)\right)=\int^{\infty }_{-\infty }{e^{-2\pi ir{\rho }_{\phi }}}\mathcal{R}\left(\mu \right)\left({\rho }_{\phi },\phi \right)d{\rho }_{\phi } \\ 
&=\int^{\infty }_{-\infty }{e^{-2\pi ir{\rho }_{\phi }}}\int^{\infty }_{-\infty }{\int^{\infty }_{-\infty }{\mu \left(x,p\right)\ }}\delta \left({\rho }_{\phi }-xcos\phi -psin\phi \right)dxdpd{\rho }_{\phi }\  \\ 
&=\int^{\infty }_{-\infty }{\int^{\infty }_{-\infty }{\mu \left(x,p\right)\ }}\left(\int^{\infty }_{-\infty }{\delta \left({\rho }_{\phi }-xcos\phi -psin\phi \right)e^{-2\pi ir{\rho }_{\phi }}d{\rho }_{\phi }}\right)dxdp\  \\ 
&=\int^{\infty }_{-\infty }{\int^{\infty }_{-\infty }{\mu \left(x,p\right)\ }}e^{-2\pi ir\left(xcos\phi +psin\phi \right)}dxdp \\ 
&=\int^{\infty }_{-\infty }{\int^{\infty }_{-\infty }{\mu \left(x,p\right)\ }}e^{-2\pi ir\left(xrcos\phi +prsin\phi \right)}dxdp. 
\end{split}
\end{array}
\end{equation} 
In \eqref{ZEqnNum935273} we exploited the shifting property of the Radon transform function, namely that
\begin{equation} \label{22)} 
 \begin{array}{l}
\begin{split}
&\mathcal{R}\left(\mu \left(\left(x,p\right)\mathrm{-}\left(b_{\mathrm{1}},b_{\mathrm{2}}\right)\right)\right) \\ 
&\mathrm{=\ }\int_{{\mathbb{R}}^{\mathrm{2}}}{\mu \left(x\mathrm{-}b_{\mathrm{1}},p\mathrm{-}b_{\mathrm{2}}\right)}\delta \left(\rho \mathrm{-}\left( \begin{array}{l}
\left(x\mathrm{-}b_{\mathrm{1}},p\mathrm{-}b_{\mathrm{2}}\right) \\ 
\mathrm{+}\left(b_{\mathrm{1}},b_{\mathrm{2}}\right) \end{array}
\right)\mathrm{\cdot }\left(\mathrm{cos}\phi \mathrm{,sin}\phi \right)\right)d\left(x\mathrm{-}b_{\mathrm{1}},p\mathrm{-}b_{\mathrm{2}}\right) \\ 
&\mathrm{=}\int_{{\mathbb{R}}^{\mathrm{2}}}{\mu \left(x\mathrm{-}b_{\mathrm{1}},p\mathrm{-}b_{\mathrm{2}}\right)}\delta \left(\rho \mathrm{-}\left( \begin{array}{l}
\left(x\mathrm{-}b_{\mathrm{1}},p\mathrm{-}b_{\mathrm{2}}\right)\mathrm{\cdot }\left(\mathrm{cos}\phi \mathrm{,sin}\phi \right) \\ 
\mathrm{+}\left(b_{\mathrm{1}},b_{\mathrm{2}}\right)\mathrm{\cdot }\left(\mathrm{cos}\phi \mathrm{,sin}\phi \right) \end{array}
\right)\right)d\left(x\mathrm{-}b_{\mathrm{1}},p\mathrm{-}b_{\mathrm{2}}\right) \\ 
&=\mathcal{R}\left(\mu \right)\left(\rho \mathrm{-}\left(b_{\mathrm{1}},b_{\mathrm{2}}\right)\mathrm{\cdot }\left(\mathrm{cos}\phi \mathrm{,sin}\phi \right),\phi \right). 
\end{split}
\end{array}
\end{equation} 
Introducing variables ${\lambda }_{\mathrm{1}}\mathrm{=}r\mathrm{cos}\phi $ and ${\lambda }_{\mathrm{2}}\mathrm{=}r\mathrm{sin}\phi $, with relations 
\begin{equation} \label{ZEqnNum925941} 
r^{\mathrm{2}}\mathrm{=}{\lambda }^{\mathrm{2}}_{\mathrm{1}}\mathrm{+}{\lambda }^{\mathrm{2}}_{\mathrm{2}} 
\end{equation} 
and
\begin{equation} \label{ZEqnNum241188} 
\mathrm{tan}\phi \mathrm{=}\frac{{\lambda }_{\mathrm{2}}}{{\lambda }_{\mathrm{1}}},             
\end{equation} 
whose connections are justified by the fact that the parameters $\left(r,\phi \right)$ play the role of polar coordinates for the plane of $\left({\lambda }_{\mathrm{1}},{\lambda }_{\mathrm{2}}\right)$, the last line of \eqref{ZEqnNum935273} can be rewritten precisely as
\begin{equation} \label{25)} 
 \begin{array}{l}
\begin{split}
F_{{\rho }_{\phi }}\left(\mathcal{R}\left({\mu }_{\phi }\right)\left(r,\phi \right)\right)&\mathrm{=\ }\int^{\mathrm{\infty }}_{\mathrm{-}\mathrm{\infty }}{\int^{\mathrm{\infty }}_{\mathrm{-}\mathrm{\infty }}{e^{\mathrm{-}\mathrm{2}\pi i\left(x{\lambda }_{\mathrm{1}}\mathrm{+}p{\lambda }_{\mathrm{2}}\right)}\mu \left(x,p\right)}}dxdp \\ 
&\mathrm{=}\int_{{\mathbb{R}}^{\mathrm{2}}}{e^{\mathrm{-}\mathrm{2}\pi i\left(x,p\right)\mathrm{\cdot }\left({\lambda }_{\mathrm{1}},{\lambda }_{\mathrm{2}}\right)}\mu \left(x,p\right)}d\left(x,p\right), 
\end{split}
\end{array}
\end{equation} 
which is, in fact, the two-dimensional Fourier transform of the partial mutual information function ${\mu }_{\phi }\left(x,p\right)\mathrm{=}I_{M_{\phi }}\left(A\mathrm{:}B\right)$.  

 The notations of our model are summarized in \fref{fig2}. All $M_{\phi ,i},i\mathrm{\in }\left[n\right]$ are modeled as abstracted lines in the high-dimensional Hilbert space. The line integral identifies an information slice that is obtained by a photon coincidence measurement at a given $\phi $. For each $\phi $, \textit{n} coincidence measurements are performed that results in the partial mutual information $I_{M_{\phi }}\left(A\mathrm{:}B\right)$, it is also referred to as ${\mu }_{\phi }\left(x,p\right)$. 

\begin{figure*}[h!]
 \begin{center}
 	 \includegraphics[angle = 0,width=1\linewidth]{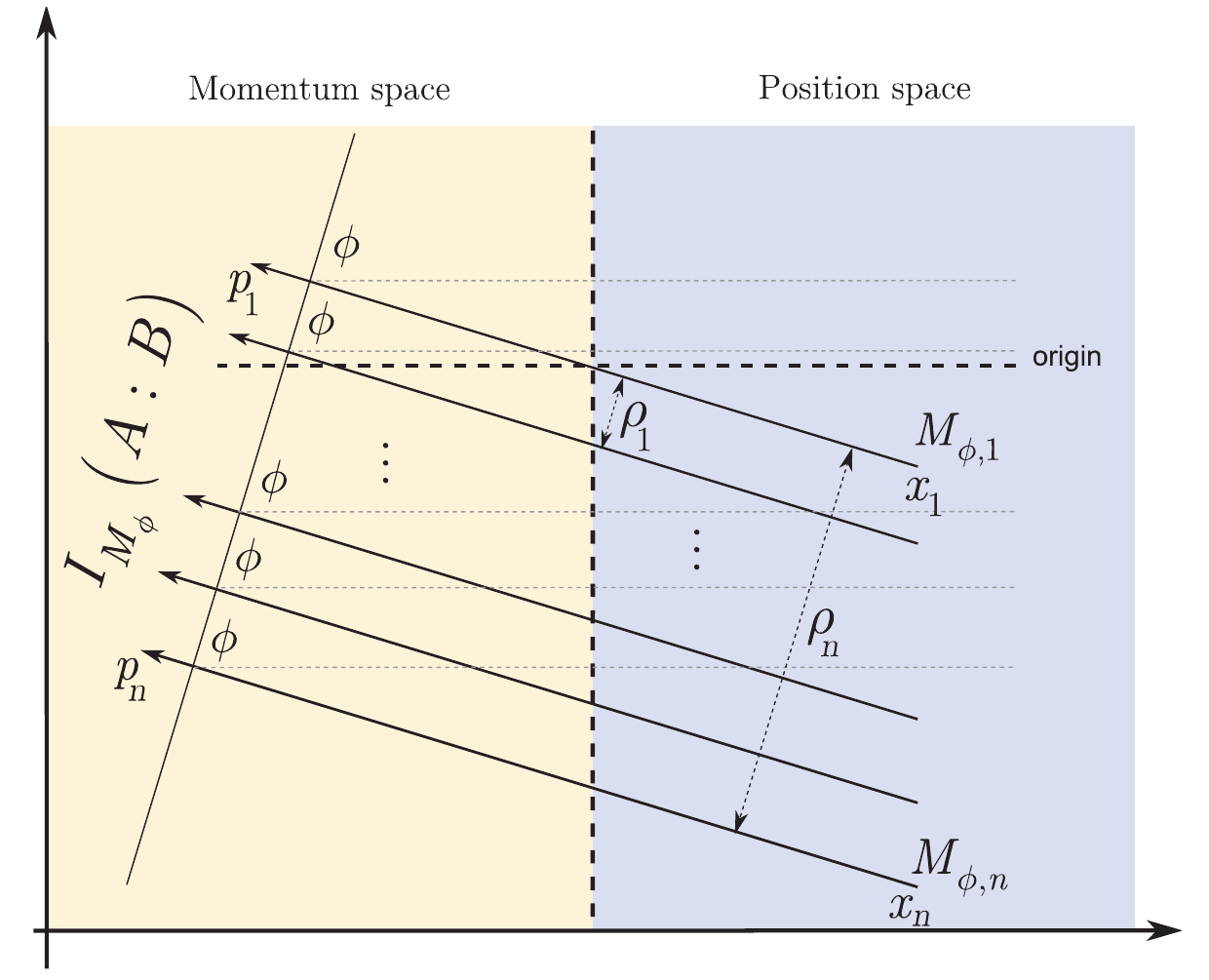}

\caption{Abstract model of the Radon transform-based quantum imaging of high-dimensional position-momentum entanglement. The integral of an abstracted line represents the slice information about the partial mutual information $I_{M_{\phi }}\left(A\mathrm{:}B\right)$. A line $M_{\phi ,i}$ represents a coincidence measurement at a given $\phi $, and the \textit{n} abstracted lines together formulate the $I_{M_{\phi }}\left(A\mathrm{:}B\right)$ partial mutual information function. Gathering together these slices for all $\phi $ indentify $I_{\mathcal{R}}\left(A\mathrm{:}B\right)$, the (full) mutual information function. (In this level of abstraction, the first abstracted line $M_{\phi \mathrm{,1}}$ assigns an origin in the position-momentum space at each $\phi $.)}
 \label{fig2}
\end{center}
\end{figure*}

 An $M_{\phi ,i}$ has a position component start point and momentum component end point, which shows that each slice covers only a piece information from the capabilities of the high-dimensional space. Each $M_{\phi }$ coincidence measurement (modeled by the abstracted lines) reveals some information about the mutual information that is carried by the high-dimensional position-momentum space. The partial mutual information function $I_{M_{\phi }}\left(A\mathrm{:}B\right)$ is defined by \textit{n} slices or abstracted lines, while the full mutual information $I_{\mathcal{R}}\left(A\mathrm{:}B\right)$ is defined by $mn$ lines, where $m$ stands for the discretized ranges of $\mathrm{0}\mathrm{\le }\phi \mathrm{<}\pi $ (for further details see \sref{sec4}). 

 For our case, the projection-slice theorem [\cref{ref30}-\cref{ref32}] in the quantum imaging of high-dimensional position-momentum entanglement can be precisely formulated as
\begin{equation} \label{ZEqnNum916578} 
F_{{\rho }_{\phi }}\left(\mathcal{R}\left({\mu }_{\phi }\right)\left(r,\phi \right)\right)\mathrm{=}F\left({\mu }_{\phi }\left({\lambda }_{\mathrm{1}},{\lambda }_{\mathrm{2}}\right)\right). 
\end{equation} 
As the next step, we rewrite \eqref{ZEqnNum916578} as 
\begin{equation} \label{ZEqnNum805246} 
F\left({\mu }_{\phi }\left({\lambda }_{\mathrm{1}},{\lambda }_{\mathrm{2}}\right)\right)\mathrm{=}\chi \left(r,\phi \right), 
\end{equation} 
where $\chi \left(r,\phi \right)$ encodes the same function as the Fourier transform of ${\mu }_{\phi }\left({\lambda }_{\mathrm{1}},{\lambda }_{\mathrm{2}}\right)$. 

 Since parameters ${\lambda }_{\mathrm{1}}$ and ${\lambda }_{\mathrm{2}}$ can be exactly computed, the unknown partial mutual information function ${\mu }_{\phi }$ can be recovered from \eqref{ZEqnNum916578} by applying the two-dimensional inverse Fourier transform. It leads to the \textit{decoded} \textit{partial mutual information function }as follows:
\begin{equation} \label{ZEqnNum542618} 
 \begin{array}{l}
\begin{split}
F^{\mathrm{-}\mathrm{1}}\left(F\left({\mu }_{\phi }\left({\lambda }_{\mathrm{1}},{\lambda }_{\mathrm{2}}\right)\right)\right)&\mathrm{=}\int_{{\mathbb{R}}^{\mathrm{2}}}{e^{\mathrm{-}\mathrm{2}\pi i\left(x,p\right)\mathrm{\cdot }\left({\lambda }_{\mathrm{1}},{\lambda }_{\mathrm{2}}\right)}F\left(\mu \left({\lambda }_{\mathrm{1}},{\lambda }_{\mathrm{2}}\right)\right)d\left({\lambda }_{\mathrm{1}},{\lambda }_{\mathrm{2}}\right)} \\ 
&\mathrm{=}{\mu }_{\phi }\left(x,p\right) \\ 
&\mathrm{=}I_{M_{\phi }}\left(A\mathrm{:}B\right). 
\end{split}
\end{array}
\end{equation} 
Applying \eqref{ZEqnNum542618} for $\mathrm{0}\mathrm{\le }\phi \mathrm{<}\pi $, the \textit{decoded full mutual information} function can be expressed as
\begin{equation} \label{29)} 
 \begin{array}{l}
\begin{split}
&F^{\mathrm{-}\mathrm{1}}\left(F\left(\int_{\phi }{{\mu }_{\phi }\left({\lambda }_{\mathrm{1}},{\lambda }_{\mathrm{2}}\right)d\phi }\right)\right)\\&\mathrm{=}\int_{\phi }{\int_{{\mathbb{R}}^{\mathrm{2}}}{e^{\mathrm{-}\mathrm{2}\pi i\left(x,p\right)\mathrm{\cdot }\left({\lambda }_{\mathrm{1}},{\lambda }_{\mathrm{2}}\right)}F\left(\mu \left({\lambda }_{\mathrm{1}},{\lambda }_{\mathrm{2}}\right)\right)d\left({\lambda }_{\mathrm{1}},{\lambda }_{\mathrm{2}}\right)d\phi }} \\ 
&\mathrm{=}\int_{\phi }{{\mu }_{\phi }\left(x,p\right)}d\phi  \\ 
&\mathrm{=}I_{\mathcal{R}}\left(A\mathrm{:}B\right), 
\end{split}
\end{array}
\end{equation} 
which concludes the proof. 
\end{proof} 
 
\subsection{Mutual Information of High-Dimensional Entanglement}
\label{sec31}
In this section we characterize the mutual information function that can be extracted from the Hilbert space by the application of the Radon transform. 

\begin{theorem}
 (Mutual information in a Radon transform.) \textit{The mutual information that can be obtained by Radon transform from a position-momentum entangled photonic system }$\left|\left.{\psi }_{AB}\right\rangle \right.$\textit{ is }$I_{\mathcal{R}}\left(A:B\right)=\int_{\phi }{I_{M_{\phi }}\left(A:B\right)}\ d\phi =\\\int_{\phi }{{\mu }_{\phi }\left(x,p\right)}\ d\phi $\textit{, where }$I_{M_{\phi }}\left(A:B\right)$\textit{ is the partial mutual information function at a given }$\phi $\textit{. }
\end{theorem}
\begin{proof}
The mutual information in the standard model can be expressed as
\begin{equation} \label{ZEqnNum372001} 
I_0\left(A\mathrm{:}B\right)\mathrm{=}I_{M_0}\left(A\mathrm{:}B\right),        
\end{equation} 
where $I_0\left(A\mathrm{:}B\right)$ stands for the mutual information that can be obtained in a standard quantum imaging scenario, at $\phi \mathrm{=0}$. Assuming a position-momentum entangled photonic bipartite system $\left|\left.{\psi }_{AB}\right\rangle \right.$, the partial mutual information function $I_{M_{\phi }}\left(A:B\right)$ at $\mathrm{0}\mathrm{\le }\phi \mathrm{<}\pi $ in the position basis is precisely evaluated as follows [\cref{ref1}, \cref{ref7}]:
\begin{equation} \label{ZEqnNum150671} 
I_{M_{\phi }}\left(A:B\right)=-\int{p\left(x_A,x_B\right)}\textnormal{log}_\textnormal{2}\left(\frac{p\left(x_A,x_B\right)}{p\left(x_A\right)p\left(x_B\right)}\right)dx_Adx_B,                     
\end{equation} 
where
\begin{equation} \label{ZEqnNum853779} 
p\left(x_A,x_B\right)\mathrm{=}{\left|f\left(x_A,x_B\right)\right|}^{\mathrm{2}}\mathrm{=}{\left|Ne^{\frac{\mathrm{-}{\left(x_A\mathrm{-}x_B\right)}^{\mathrm{2}}}{\mathrm{4}w^{\mathrm{2}}_{\mathrm{1}}}}e^{\frac{\mathrm{-}{\left(x_A\mathrm{+}x_B\right)}^{\mathrm{2}}}{\mathrm{16}w^{\mathrm{2}}_{\mathrm{2}}}}\right|}^{\mathrm{2}}, 
\end{equation} 
and
\begin{equation} \label{33)} 
p\left(x_A\right)\mathrm{=}\int{{\left|f\left(x_A,x_B\right)\right|}^{\mathrm{2}}}dx_B\mathrm{=}{\left|Ne^{\frac{\mathrm{-}{\left(x_A\mathrm{-}x_B\right)}^{\mathrm{2}}}{\mathrm{4}w^{\mathrm{2}}_{\mathrm{1}}}}e^{\frac{\mathrm{-}{\left(x_A\mathrm{+}x_B\right)}^{\mathrm{2}}}{\mathrm{16}w^{\mathrm{2}}_{\mathrm{2}}}}\right|}^{\mathrm{2}}dx_B, 
\end{equation} 
while
\begin{equation} \label{ZEqnNum536377} 
p\left(x_B\right)\mathrm{=}\int{{\left|f\left(x_A,x_B\right)\right|}^{\mathrm{2}}}dx_A\mathrm{=}{\left|Ne^{\frac{\mathrm{-}{\left(x_A\mathrm{-}x_B\right)}^{\mathrm{2}}}{\mathrm{4}w^{\mathrm{2}}_{\mathrm{1}}}}e^{\frac{\mathrm{-}{\left(x_A\mathrm{+}x_B\right)}^{\mathrm{2}}}{\mathrm{16}w^{\mathrm{2}}_{\mathrm{2}}}}\right|}^{\mathrm{2}}dx_A, 
\end{equation} 
and where 
\begin{equation} \label{35)} 
N\mathrm{=}\frac{\mathrm{1}}{\mathrm{2}\pi w_{\mathrm{1}}w_{\mathrm{2}}}.                                                    
\end{equation} 
The same relations hold for the momentum basis (for simplicity we do not repeat here these equations), i.e.,
\begin{equation} \label{ZEqnNum961261} 
I_{M_{\phi }}\left(A:B\right)=-\int{p\left(p_A,p_B\right)}\textnormal{log}_\textnormal{2}\left(\frac{p\left(p_A,p_B\right)}{p\left(p_A\right)p\left(p_B\right)}\right)dp_Adp_B.                     
\end{equation} 
One can readily see that \eqref{ZEqnNum961261} is, in fact, the mutual information that can be extracted from a standard coincidence detection at a fixed $\phi $ -- i.e., the partial mutual information function (\textit{Note}: ``partial'', in terms of the Radon transform-based approach means ``full'' in terms of the standard model at $\phi \mathrm{=0}$) represents the full mutual information that can be reached in a standard quantum imaging scenario. The reason behind this: First, in the standard model, there is no phase modulation $\phi $ in any path, which allows no distinction to be made with respect to $\phi $ in the data processing. Second, the Radon transform-based post-processing allows one to extract more information from these information slices than the standard scenario without Radon transform.  

 From \eqref{ZEqnNum853779} and \eqref{ZEqnNum961261}, the partial mutual information function $I_{M_{\phi }}\left(A:B\right)$ of a position-momentum entangled photonic system $\left|\left.{\psi }_{AB}\right\rangle \right.$ is evaluated as
\begin{equation} \label{ZEqnNum873514} 
I_{M_{\phi }}\left(A:B\right)=\textnormal{log}_\textnormal{2}{\left(\frac{{\sigma }_s}{{\sigma }_C}\right)}^2=\textnormal{log}_\textnormal{2}{\left(\frac{4w^2_2+w^2_1}{4w^2_1w^2_2}\right)}^2, 
\end{equation} 
where the noise parameter ${\sigma }_C$ (see \eqref{ZEqnNum574414} and \eqref{ZEqnNum516221}) is not an additive noise, in contrast to the standard Shannon model [\cref{ref1}, \cref{ref7}]. However, by introducing 
\begin{equation} \label{38)} 
N\mathrm{=}\frac{{\sigma }^{\mathrm{2}}_s}{\mathrm{1-}\frac{{\sigma }^{\mathrm{2}}_C}{{\sigma }^{\mathrm{2}}_s}}\mathrm{\approx }{\sigma }_C,                                             
\end{equation} 
which is satisfied if ${\sigma }_C\mathrm{\ll }{\sigma }_s$, the result is a purely additive noise [\cref{ref1}, \cref{ref7}], i.e., the partial information function picks up the standard Shannon formula
\begin{equation} \label{39)} 
I_{M_{\phi }}\left(A:B\right)=\textnormal{log}_\textnormal{2}\left(1+\frac{{\sigma }^2_s}{N^2}\right). 
\end{equation} 
In the Radon transform-based setting, the mutual information function of a position-momentum entangled photonic system $\left|\left.{\psi }_{AB}\right\rangle \right.$ for $\mathrm{0}\mathrm{\le }\phi \mathrm{<}\pi $ is 
\begin{equation} \label{ZEqnNum737158} 
 \begin{array}{l}
\begin{split}
I_{\mathcal{R}}\left(A\mathrm{:}B\right)&\mathrm{=}\int_{\phi }{I_{M_{\phi }}\left(A\mathrm{:}B\right)}d\phi  \\ 
&\mathrm{=-}\int_{\phi }{\int{p\left(x_A,x_B\right)}\mathrm{lo}{\mathrm{g}}_{\mathrm{2}}\left(\frac{p\left(x_A,x_B\right)}{p\left(x_A\right)p\left(x_B\right)}\right)dx_Adx_B}d\phi  \\ 
&\mathrm{=}\int_{\phi }{{\mu }_{\phi }\left(x,p\right)}d\phi , 
\end{split}
\end{array}
\end{equation} 
which is the theoretical maximum that can be reached in our setting, not taking into account the parameters of the measurement apparatus.

 In terms of coincidence detection events $M^A_{\phi ,i}$ and $M^B_{\phi ,i}$ at the two paths \textit{A} and \textit{B}, at a given $\phi $ the correlations can be exactly quantified. Let $M^A_{\phi }\mathrm{=}\sum_a{M^A_{\phi ,i}}$ and $M^B_{\phi }\mathrm{=}\sum_b{M^B_{\phi ,i}}$, $a,b\mathrm{>0}$, $a\mathrm{\ne }b$. 

 Then, for the position basis the joint detection probability is as
\begin{equation} \label{41)} 
\mathrm{Pr}\left(M^A_{\phi },M^B_{\phi }\right)\mathrm{=}\int_{M^A_{\phi }}{dx_A}\int_{M^B_{\phi }}{dx_B}{\left|f\left(x_A,x_B\right)\right|}^{\mathrm{2}},                          
\end{equation} 
and for the momentum basis it is precisely evaluated as
\begin{equation} \label{42)} 
\mathrm{Pr}\left(M^A_{\phi },M^B_{\phi }\right)\mathrm{=}\int_{M^A_{\phi }}{dp_A}\int_{M^B_{\phi }}{dp_B}{\left|f\left(p_A,p_B\right)\right|}^{\mathrm{2}},                           
\end{equation} 
hence the partial mutual information function can be expressed as
\begin{equation} \label{ZEqnNum947675} 
 \begin{array}{l}
\begin{split}
I_{M_{\phi }}\left(A:B\right)=\sum_{M^A_{\phi }}&{Pr\left(M^A_{\phi }\right)}\textnormal{log}_\textnormal{2}Pr\left(M^A_{\phi }\right)+\sum_{M^B_{\phi }}{Pr\left(M^B_{\phi }\right)}\textnormal{log}_\textnormal{2}Pr\left(M^B_{\phi }\right) \\& 
-\sum_{M^A_{\phi },M^B_{\phi }}{Pr\left(M^A_{\phi },M^B_{\phi }\right)}\textnormal{log}_\textnormal{2}Pr\left(M^A_{\phi },M^B_{\phi }\right), 
\end{split}
\end{array}
\end{equation} 
where
\begin{equation} \label{44)} 
\mathrm{Pr}\left(M^A_{\phi }\right)\mathrm{=}\sum_{M^B_{\phi }}{\mathrm{Pr}\left(M^A_{\phi },M^B_{\phi }\right)},                                     
\end{equation} 
and
\begin{equation} \label{45)} 
\mathrm{Pr}\left(M^B_{\phi }\right)\mathrm{=}\sum_{M^A_{\phi }}{\mathrm{Pr}\left(M^A_{\phi },M^B_{\phi }\right)}.                                     
\end{equation} 
In terms of the joint detection probability, the mutual information under the Radon transform leads to
\begin{equation} \label{ZEqnNum119461} 
 \begin{array}{l}
\begin{split}
I_{\mathcal{R}}\left(A:B\right)=\sum_{\phi }&{\left(\sum_{M^A_{\phi }}{Pr\left(M^A_{\phi }\right)}\right.\textnormal{log}_\textnormal{2}Pr\left(M^A_{\phi }\right)}+\sum_{M^B_{\phi }}{Pr\left(M^B_{\phi }\right)}\textnormal{log}_\textnormal{2}Pr\left(M^B_{\phi }\right) \\& 
-\sum_{M^A_{\phi },M^B_{\phi }}{Pr\left(M^A_{\phi },M^B_{\phi }\right)}\textnormal{log}_\textnormal{2}\left.Pr\left(M^A_{\phi },M^B_{\phi }\right)\right), 
\end{split}
\end{array}
\end{equation} 
which indeed quantifies the same amount of information as \eqref{ZEqnNum737158}, which concludes the proof. 
\end{proof}

\subsection{Stronger Violation of Entropic Separability Bound}
In this section we reveal that under Radon transform, the separability bound is violated more significantly than in a standard quantum imaging scenario, which indicates lower conditional entropy and the higher information transmission capability of a photonic position-momentum entangled state. 

\begin{theorem}
The $I_{\mathcal{R}}\left(A:B\right)$\textit{ mutual information obtained from the Radon transform is closer to the theoretical maximum }$\textnormal{max}_{\forall {\rho }_i}I\left(A:B\right)=\textnormal{log}_\textnormal{2}\left(d\right)$\textit{ than the mutual information }$I_0\left(A:B\right)$\textit{ of the standard photon coincidence detection, where }$d$\textit{ is the dimension of the detector. At a given d, the entropic separability bound is violated by more standard deviations in comparison to the standard model.}
\end{theorem}
\begin{proof}
 According to the entropic separability bound (SB), the measurement data obtained in the \textit{x} position and \textit{p} momentum bases can be used to quantify the information transmission capability of a photonic entangled state. 
 The conditional entropy $H\left(\left.A\right|B\right)\mathrm{=}H\left(AB\right)\mathrm{-}H\left(B\right)$ in terms of position and momentum measurements can be rephrased as
\begin{equation} \label{ZEqnNum353900} 
H{\left(\left.A\right|B\right)}_x\mathrm{+}H{\left(\left.A\right|B\right)}_p\mathrm{=}\left(H{\left(AB\right)}_x\mathrm{-}H{\left(B\right)}_x\right)\mathrm{+}\left(H{\left(AB\right)}_p\mathrm{-}H{\left(B\right)}_p\right).         
\end{equation} 
From \eqref{ZEqnNum353900}, the entropic separability bound can be stated as follows. Any bipartite position-momentum quantum system can be entangled if only the following inequality of the entropic uncertainty holds [\cref{ref1}, \cref{ref7}]:
\begin{equation} \label{48)} 
\boldsymbol{\mathrm{SB}}\mathrm{:}H{\left(\left.A\right|B\right)}_x\mathrm{+}H{\left(\left.A\right|B\right)}_p\mathrm{<6.18}.                 
\end{equation} 
As the detector dimension \textit{d} increases, one can readily find the following. The sum of conditional entropies $H{\left(\left.A\right|B\right)}_x\mathrm{+}H{\left(\left.A\right|B\right)}_p$, $H{\left(\left.B\right|A\right)}_x\mathrm{+}H{\left(\left.B\right|A\right)}_p$ starts to converge to zero, which means that as \textit{d} increases a deeper and more appropriate description of the information conveying the capability of the position-momentum entangled system becomes available. The depth of the quantum imaging process is limited by the measurement apparatus, i.e., the amount of detectable mutual information that is also upper bounded by precisely $\textnormal{max}_{\forall {\rho }_i}I\left(A:B\right)=\textnormal{log}_\textnormal{2}\left(d\right)$.

 In a standard model, at a given \textit{d}, the violation (denoted by $\boldsymbol{\mathrm{S}}{\boldsymbol{\mathrm{B}}}_0$) can be rephrased in terms of the $\sigma $ standard deviation, as follows [\cref{ref1}, \cref{ref7}]:
\begin{equation} \label{ZEqnNum987686} 
\boldsymbol{\mathrm{S}}{\boldsymbol{\mathrm{B}}}_0\mathrm{=}H{\left(\left.A\right|B\right)}_x\mathrm{+}H{\left(\left.A\right|B\right)}_p\mathrm{=}\boldsymbol{\mathrm{SB}}\mathrm{\cdot }\frac{\mathrm{1}}{\tau \sigma },         
\end{equation} 
and 
\begin{equation} \label{ZEqnNum972118} 
\boldsymbol{\mathrm{S}}{\boldsymbol{\mathrm{B}}}_0\mathrm{=}H{\left(\left.B\right|A\right)}_x\mathrm{+}H{\left(\left.B\right|A\right)}_p\mathrm{=}\boldsymbol{\mathrm{SB}}\mathrm{\cdot }\frac{\mathrm{1}}{\tau \sigma }.         
\end{equation} 
If \eqref{ZEqnNum987686} and \eqref{ZEqnNum972118} hold, then the separability bound is violated by $\tau $ standard deviations.  

 Assuming that in the standard model the violation is $\tau $, for the Radon transform-based quantum imaging from \eqref{ZEqnNum119461} the violation is $\kappa $ standard deviations, as
\begin{equation} \label{51)} 
\kappa \mathrm{>}\tau ,                      
\end{equation} 
and for the entropic separability bound (denoted by $\boldsymbol{\mathrm{S}}{\boldsymbol{\mathrm{B}}}_{\mathcal{R}}$) one obtains
\begin{equation} \label{52)} 
\boldsymbol{\mathrm{S}}{\boldsymbol{\mathrm{B}}}_{\mathcal{R}}\mathrm{=}H{\left(\left.A\right|B\right)}_x\mathrm{+}H{\left(\left.A\right|B\right)}_p\mathrm{=}\boldsymbol{\mathrm{SB}}\mathrm{\cdot }\frac{\mathrm{1}}{\kappa \sigma }\mathrm{<}\boldsymbol{\mathrm{S}}{\boldsymbol{\mathrm{B}}}_0 
\end{equation} 
and
\begin{equation} \label{53)} 
\boldsymbol{\mathrm{S}}{\boldsymbol{\mathrm{B}}}_{\mathcal{R}}\mathrm{=}H{\left(\left.B\right|A\right)}_x\mathrm{+}H{\left(\left.B\right|A\right)}_p\mathrm{=}\boldsymbol{\mathrm{SB}}\mathrm{\cdot }\frac{\mathrm{1}}{\kappa \sigma }\mathrm{<}\boldsymbol{\mathrm{S}}{\boldsymbol{\mathrm{B}}}_0,                        
\end{equation} 
hence, the violation of the separability bound is stronger at the given dimension. This result indicates more significantly the presence of quantum influences than the standard model, and also reveals that the analyzed space cannot be simulated (replicated) in a classical framework. These statements are summarized in \fref{fig3}. 

\begin{figure*}[h!]
 \begin{center}
 	 \includegraphics[angle = 0,width=1\linewidth]{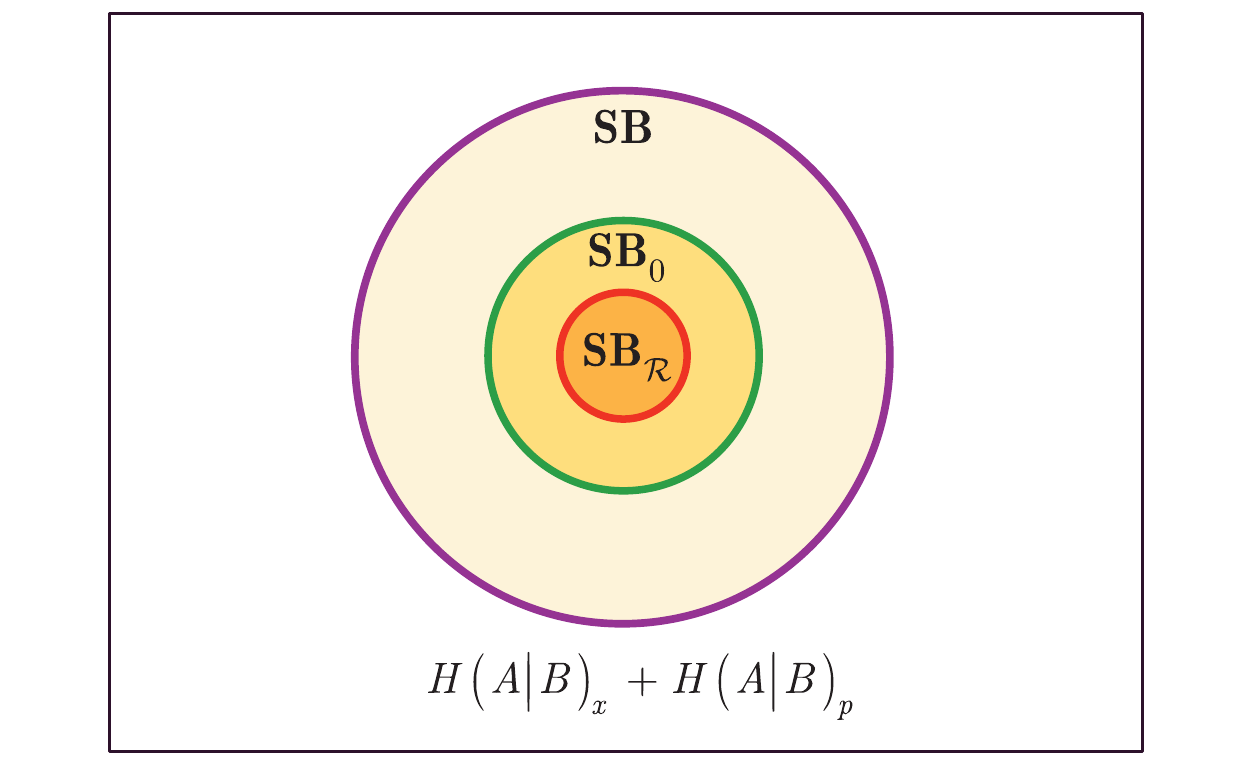}

\caption{The violation of the separability bound ($\boldsymbol{\mathrm{SB}}$). In the standard model ($\boldsymbol{\mathrm{S}}{\boldsymbol{\mathrm{B}}}_0$), the violation is $\tau $ standard deviations. In the Radon transform-based setting ($\boldsymbol{\mathrm{S}}{\boldsymbol{\mathrm{B}}}_{\mathcal{R}}$), the violation is stronger, $\kappa \mathrm{>}\tau $ standard deviations.}
 \label{fig3}
\end{center}
\end{figure*}

 To demonstrate these statements, we present a numerical analysis. We use the system parameterization of [\cref{ref1}, \cref{ref7}], i.e., the position-momentum entanglement is characterized as follows. The input laser source has a wavelength of 325 nm, and ${\sigma }_s\mathrm{=1500}$ $\mu$m and ${\sigma }_C\mathrm{=40}$ $\mu$m. Based on these parameters, the optimal mutual information function $I_0\left(A\mathrm{:}B\right)$ of the standard model in the position basis at $\phi \mathrm{=0}$ can be exactly evaluated by the joint detection events \eqref{ZEqnNum947675} at $d\mathrm{\to }\mathrm{\infty }$ as $I_0\left(A:B\right)\approx$ 10 bits/photon. 

 In the Radon transform the optimum is different; the correct formula at $d\mathrm{\to }\mathrm{\infty }$ is \eqref{ZEqnNum119461}, which leads to $I_{\mathcal{R}}\left(A\mathrm{:}B\right)\mathrm{\approx }\mathrm{13}$ ${\mathrm{bits}}/{\mathrm{photon}}$, for the range of $\mathrm{0}\mathrm{\le }\phi \mathrm{<}\pi $. Hence, the optimal amount of the extractable mutual information can be increased in the asymptotic limit of the measurement space. The same connections hold for the momentum basis. 

 Using this input system parameterization, the mutual information of \eqref{ZEqnNum947675} and \eqref{ZEqnNum119461} are shown in \fref{fig4}. Using the position basis, these quantities are first depicted in \fref{fig4}(a) for a fixed dimension, $d\mathrm{=900}$. In \fref{fig4}(b), the quantities are depicted as a function of the dimension, $\mathrm{0}\mathrm{\le }d\mathrm{\le }\mathrm{1000}$. 

\begin{figure*}[h!]
 \begin{center}
 	 \includegraphics[angle = 0,width=1\linewidth]{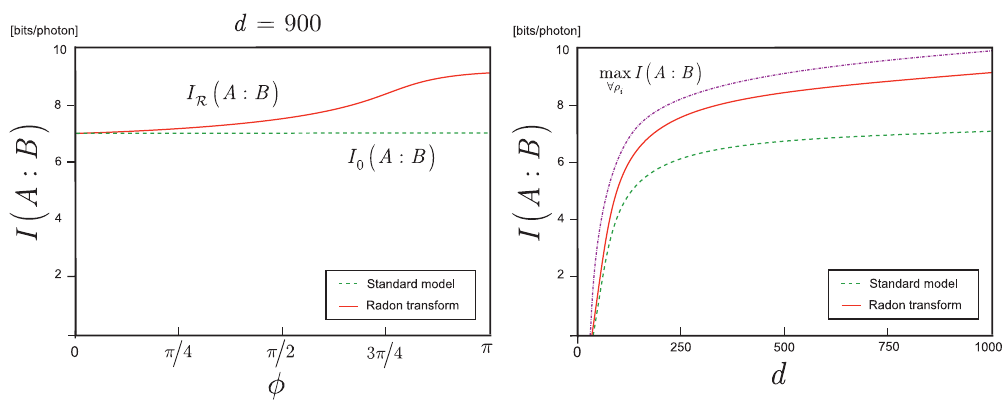}

\caption{(a): The partial mutual information (dashed green) and the full mutual information obtained by Radon transform (single red) at a fixed dimension. (b): The mutual information of the standard model and the Radon transform as a function of the dimension. The theoretical maximum, $\textnormal{log}_\textnormal{2}\left(d\right)$,\textit{ }is depicted by the dash-dotted purple line. The curves are obtained from \eqref{ZEqnNum947675} and \eqref{ZEqnNum119461}, respectively.}
 \label{fig4}
\end{center}
\end{figure*}

The results show that the Radon transform-based model offers higher extractable mutual information at an arbitrary dimension. The quantity $I_{\mathcal{R}}\left(A\mathrm{:}B\right)$ of the Radon transform approximates more precisely the theoretical upper bound $\textnormal{log}_\textnormal{2}\left(d\right)$ than the mutual information of the standard model $I_0\left(A:B\right)$. The Radon transform-based measurement setup enhances the accuracy of the tomography process, and reveals those hidden fractions that are not sampled and are not processed in the standard model. These results conclude the proof. 
\end{proof}

The analysis revealed that for any \textit{d}, the mutual information obtained in the Radon transform-based model is closer to the theoretical maximum than that of the standard model. 
 
\subsection{Application in Continuous-Variable Quantum Key Distribution}

 In Continuous-Variable Quantum Key Distribution (CVQKD), the information is conveyed by Gaussian random distributed coherent states. Let $\psi \left(x,p\right)$ be a Gaussian random state in the phase space 
\begin{equation} \label{ZEqnNum561886} 
\psi \left(x,p\right)\mathrm{=}\frac{\mathrm{1}}{\mathrm{2}\pi {\sigma }^{\mathrm{2}}}e^{\frac{\mathrm{-}\left(x^{\mathrm{2}}\mathrm{+}p^{\mathrm{2}}\right)}{\mathrm{2}{\sigma }^{\mathrm{2}}}}, 
\end{equation} 
with zero mean, i.i.d. Gaussian random position and momentum quadratures $x,p\in {\rm {\mathcal N}}\left(0,\sigma ^{2} \right)$. 

 The Radon transform for this Gaussian random distribution can be calculated as follows:
\begin{equation} \label{ZEqnNum883793} 
\mathcal{R}\left(\mu \left(\rho ,\phi \right)\right)\mathrm{=}\int^{\mathrm{\infty }}_{\mathrm{-}\mathrm{\infty }}{\int^{\mathrm{\infty }}_{\mathrm{-}\mathrm{\infty }}{\frac{\mathrm{1}}{\mathrm{2}\pi {\sigma }^{\mathrm{2}}}e^{\frac{\mathrm{-}\left(x^{\mathrm{2}}\mathrm{+}p^{\mathrm{2}}\right)}{\mathrm{2}{\sigma }^{\mathrm{2}}}}}}\delta \left(\rho \mathrm{-}x\mathrm{cos}\phi \mathrm{-}p\mathrm{sin}\phi \right)dxdp. 
\end{equation} 
Introducing $u_{\mathrm{1}}\mathrm{=}x\mathrm{cos}\phi \mathrm{+}p\mathrm{si}\mathrm{n}\phi $, $u_{\mathrm{2}}\mathrm{=-}x\mathrm{sin}\phi \mathrm{+}p\mathrm{cos}\phi $, with $u^{\mathrm{2}}_{\mathrm{1}}\mathrm{+}u^{\mathrm{2}}_{\mathrm{2}}\mathrm{=}x^{\mathrm{2}}\mathrm{+}p^{\mathrm{2}}$ [\cref{ref30}-\cref{ref32}], \eqref{ZEqnNum883793} can be rewritten as
\begin{equation} \label{56)} 
 \begin{array}{l}
\begin{split}
\mathcal{R}\left(\mu \left(\rho ,\phi \right)\right)&\mathrm{=}\int^{\mathrm{\infty }}_{\mathrm{-}\mathrm{\infty }}{\int^{\mathrm{\infty }}_{\mathrm{-}\mathrm{\infty }}{\frac{\mathrm{1}}{\mathrm{2}\pi {\sigma }^{\mathrm{2}}}e^{\frac{\mathrm{-}\left(u^{\mathrm{2}}_{\mathrm{1}}\mathrm{+}u^{\mathrm{2}}_{\mathrm{2}}\right)}{\mathrm{2}{\sigma }^{\mathrm{2}}}}}}\delta \left(\rho \mathrm{-}u_{\mathrm{1}}\right)du_{\mathrm{1}}du_{\mathrm{2}} \\ 
&\mathrm{=}\int^{\mathrm{\infty }}_{\mathrm{-}\mathrm{\infty }}{\frac{\mathrm{1}}{\mathrm{2}\pi {\sigma }^{\mathrm{2}}}\left(\int^{\mathrm{\infty }}_{\mathrm{-}\mathrm{\infty }}{\frac{\mathrm{1}}{\mathrm{2}\pi {\sigma }^{\mathrm{2}}}e^{\frac{\mathrm{-}u^{\mathrm{2}}_{\mathrm{1}}}{\mathrm{2}{\sigma }^{\mathrm{2}}}}\delta \left(\rho \mathrm{-}u_{\mathrm{1}}\right)du_{\mathrm{1}}}\right)}e^{\frac{\mathrm{-}u^{\mathrm{2}}_{\mathrm{2}}}{\mathrm{2}{\sigma }^{\mathrm{2}}}}du_{\mathrm{2}} \\ 
&\mathrm{=}\int^{\mathrm{\infty }}_{\mathrm{-}\mathrm{\infty }}{\frac{\mathrm{1}}{\mathrm{4}\pi {\sigma }^{\mathrm{4}}}e^{\frac{\mathrm{-}{\rho }^{\mathrm{2}}}{\mathrm{2}{\sigma }^{\mathrm{2}}}}}e^{\frac{\mathrm{-}u^{\mathrm{2}}_{\mathrm{2}}}{\mathrm{2}{\sigma }^{\mathrm{2}}}}du_{\mathrm{2}} \\ 
&\mathrm{=}\frac{\mathrm{1}}{\mathrm{4}\pi {\sigma }^{\mathrm{4}}}e^{\frac{\mathrm{-}{\rho }^{\mathrm{2}}}{\mathrm{2}{\sigma }^{\mathrm{2}}}}\int^{\mathrm{\infty }}_{\mathrm{-}\mathrm{\infty }}{e^{\frac{\mathrm{-}u^{\mathrm{2}}_{\mathrm{2}}}{\mathrm{2}{\sigma }^{\mathrm{2}}}}}du_{\mathrm{2}} \\ 
&\mathrm{=}\frac{\mathrm{1}}{\mathrm{4}\pi {\sigma }^{\mathrm{4}}}e^{\frac{\mathrm{-}{\rho }^{\mathrm{2}}}{\mathrm{2}{\sigma }^{\mathrm{2}}}}, 
\end{split}
\end{array}
\end{equation} 
where the last line is justified by the normalization of the Gaussian [\cref{ref30}-\cref{ref32}]. 

 Using $r\mathrm{=}x^{\mathrm{2}}\mathrm{+}p^{\mathrm{2}}$, one gets the polar form of the Radon-transformed Gaussian as
\begin{equation} \label{57)} 
\mathcal{R}\left(\mu \left(\rho ,\phi \right)\right)=\mathcal{R}\left(\frac{\mathrm{1}}{\mathrm{2}\pi {\sigma }^{\mathrm{2}}}e^{\frac{\mathrm{-}r}{\mathrm{2}{\sigma }^{\mathrm{2}}}}\right)\mathrm{=}\frac{\mathrm{1}}{\mathrm{4}\pi {\sigma }^{\mathrm{4}}}e^{\frac{\mathrm{-}{\rho }^{\mathrm{2}}}{\mathrm{2}{\sigma }^{\mathrm{2}}}},             
\end{equation} 
where $\rho \mathrm{=}x\mathrm{cos}\phi \mathrm{+}p\mathrm{sin}\phi $.
 
\section{Numerical Post-Processing}
\label{sec4}
 In this section we reformulate the Radon transform-based quantum imaging in the language of data processing and interpret it as a numerical post-processing task.
\begin{lemma}
Radon transform-based quantum imaging can be implemented by numerical post-processing on the raw data.
\end{lemma}
\begin{proof}
 In \eqref{ZEqnNum925941} and \eqref{ZEqnNum241188} we have seen that parameters $\left(r,\phi \right)$ can be viewed as polar coordinates for the $\left({\lambda }_{\mathrm{1}},{\lambda }_{\mathrm{2}}\right)$ plane. We step forward from this point. Let assume that from the photon coincidence detections, the encoded partial mutual information function $\mathcal{E}\left({\mu }_{\phi }\left({\rho }_{\phi },\phi \right)\right)$ is obtained. First, the values of $\phi $ are discretized as ${\phi }_j\mathrm{=}{j\pi }/{m}$, $j\mathrm{=1,\dots ,}m$. After a normalization of $\mathrm{0}\mathrm{\le }{\rho }_{\phi }\mathrm{<1}$, the $\left({\lambda }_{\mathrm{1}},{\lambda }_{\mathrm{2}}\right)$ plane can be restricted to the complex unit circle, and if $n$ coincidence measurements are performed for each ${\phi }_j$, then ${\rho }_l\mathrm{=}{l}/{n},$ $l\mathrm{=1\dots }n$, which represent $mn$ measurements in overall. 

 Let ${\mu }^{lj}_{mn}\mathrm{=}\mu \left({\rho }_l,{\phi }_j\right)$, then the resulting function can be expressed as
\begin{equation} \label{58)} 
{\mu }_{mn}\mathrm{=}\sum^n_{l\mathrm{=1}}{\sum^m_{j\mathrm{=1}}{{\mu }^{lj}_{mn}}}\mathrm{=}I_{\mathcal{R}}\left(A\mathrm{:}B\right),                            
\end{equation} 
i.e., it contains all information from the mutual information function.

 The frequency variable $r$ can also be discretized. From the sampling theorem follows that in the computation of $F\left({\mu }^{lj}_{mn}\left(r,{\phi }_j\right)\right)$ parameter \textit{r} has to be parameterized as $r_f\mathrm{=}{f}/{\mathrm{2}}$, $f\mathrm{=1\dots }n$. 

 Applying the results of Theorem 1, one obtains
\begin{equation} \label{ZEqnNum420108} 
F\left({\mu }^{lj}_{mn}\left(r,{\phi }_j\right)\right)\mathrm{=}\int^{\mathrm{1}}_0{e^{\mathrm{-}\mathrm{2}\pi ir{\rho }_{\phi }}}{\mu }^{lj}_{mn}d{\rho }_{\phi }.                      
\end{equation} 
This function can be easily evaluated at ${\rho }_l\mathrm{=}{l}/{n}$ by the trapezoidal rule [\cref{ref30}-\cref{ref32}]. Using ${\rho }_l$$\mathrm{=}{l}/{n}$, ${\phi }_j\mathrm{=}{j\pi }/{m}$, and $r_f\mathrm{=}{f}/{\mathrm{2}}$, at a given \textit{f}  the Fourier transform is evaluated as
\begin{equation} \label{ZEqnNum783664} 
 \begin{array}{l}
\begin{split}
F\left(\mu \left(r_f,{\phi }_j\right)\right)&\mathrm{=2}\mathrm{\cdot }\frac{\mathrm{1}}{n}\sum_l{e^{\mathrm{-}\mathrm{2}\pi ir_f{\rho }_l}{\mu }^{lj}_{mn}}\left({\rho }_l,{\phi }_j\right) \\ 
&\mathrm{=2}\mathrm{\cdot }\frac{\mathrm{1}}{n}\sum_l{e^{\frac{\mathrm{-}\mathrm{2}\pi iml}{n}}{\mu }^{lj}_{mn}\left({\rho }_l,{\phi }_j\right)} \\ 
&\mathrm{=2}\mathrm{\cdot }\frac{\mathrm{1}}{n}F\left({\mu }^{lj}_{mn}\left({\rho }_l,{\phi }_j\right)\left[f\right]\right), 
\end{split}
\end{array}
\end{equation} 
where factor 2 is a corollary from the trapezoidal rule. 

 From \eqref{ZEqnNum783664}, a given slice identified with indices \textit{i},\textit{j} can be rewritten as
\begin{equation} \label{ZEqnNum658793} 
\chi \left(r_f,{\phi }_j\right)\mathrm{=}F\left({\mu }^{lj}_{mn}\left({\rho }_l,{\phi }_j\right)\left[f\right]\right),           
\end{equation} 
while computing \eqref{ZEqnNum783664} for all ${\rho }_l\mathrm{=}{l}/{n},$ $l\mathrm{=1\dots }n$ at a fixed ${\phi }_j$ results in \eqref{ZEqnNum805246}.

 The information that is contained in $\chi \left(r_f,{\phi }_j\right)$ can be represented by a polar coordinate grid in the frequency domain, and each $\chi \left(r_f,{\phi }_j\right)$ is a data point in the grid. 

 The polar grid point can be rewritten as Cartesian grid points by using the weighted average of the polar grid points [\cref{ref30}-\cref{ref32}], as 
\begin{equation} \label{62)} 
\mathcal{C}\left(\chi \right)\mathrm{=}w_{\mathrm{1}}\left({\chi }_{\mathrm{1}}\right)\mathrm{+}w_{\mathrm{2}}\left({\chi }_{\mathrm{2}}\right)\mathrm{+}w_{\mathrm{3}}\left({\chi }_{\mathrm{3}}\right)\mathrm{+}w_{\mathrm{4}}\left({\chi }_{\mathrm{4}}\right),     
\end{equation} 
where ${\chi }_i$ are the nearest neighbors, while $w_i$ are the weights of the polar grid data points.

 In terms of the Cartesian data points, the function of \eqref{ZEqnNum658793} that is resulted from the photon coincidence detections leads to the inverse Fourier transformed Cartesian 
\begin{equation} \label{63)} 
F^{\mathrm{-}\mathrm{1}}\left(\mathcal{C}\left(\chi \right)\right),              
\end{equation} 
which, in fact, encodes an information slice (see \eqref{ZEqnNum264887}) of the partial mutual information function $I_{M_{\phi }}\left(A\mathrm{:}B\right)$. 

 Extending the process for $l\mathrm{=1\dots }n$, $j\mathrm{=1,\dots ,}m$, and $f\mathrm{=1\dots }n$, the full mutual information from the Cartesian data points can be recovered as
\begin{equation} \label{64)} 
 \begin{array}{l}
\begin{split}
&\sum_f{\sum_j{F^{\mathrm{-}\mathrm{1}}\left(\mathcal{C}\left(\chi \left(r_f,{\phi }_j\right)\right)\right)}}\\ 
&\mathrm{=}\sum_l{\sum_j{\sum_f{F^{\mathrm{-}\mathrm{1}}\left(\mathcal{C}\left(F\left({\mu }^{lj}_{mn}\left({\rho }_l,{\phi }_j\right)\left[f\right]\right)\right)\right)}}} \\ 
&\mathrm{=}\sum_m{I_{M_{\phi }}\left(A\mathrm{:}B\right)} \\ 
&\mathrm{=}I_{\mathcal{R}}\left(A\mathrm{:}B\right). 
\end{split}
\end{array}
\end{equation}
The proof is therefore concluded here. 
\end{proof}
\section{Conclusions}
\label{sec5}
The exact characterization of the information coding and transmission capabilities that lie in high-dimensional Hilbert spaces is a crucial cornerstone from the viewpoint of the performance analysis of quantum communication protocols. Quantum entanglement has several important consequences in practical engineering. In particular, the high-dimensional entangled quantum systems offer several advantages and benefits in communication scenarios, and represent an essential ingredient in high-performance quantum protocols. Since the possibilities in the physical layer manipulations of quantum imaging are strongly limited, we had to find a different answer for the sharpening. We introduced a Radon transform-based quantum imaging technique for high-dimensional Hilbert spaces, which uses the raw data of the measurements and a carefully constructed post-processing for the enhancing. We showed that the theoretical upper bound of maximally extractable mutual information can be approached more closely, which allows a clearer and sharper image to be drawn from the information transmission capabilities of high-dimensional Hilbert spaces. We also revealed that the Radon transform-based quantum imaging violates much more significantly the entropic separability bound than the standard model, which indicates the presence of stronger quantum influences.
 
\section*{Acknowledgements}
This work was partially supported by the GOP-1.1.1-11-2012-0092 project sponsored by the EU and European Structural Fund, by the Hungarian Scientific Research Fund - OTKA K-112125, and by the COST Action MP1006. 

\section*{References}
\begin{enumerate}[ {[}1{]} ]
\item \label{ref1} P. Ben Dixon, Gregory A. Howland, James Schneeloch, and John C. Howell, Quantum Mutual Information Capacity for High-dimensional Entangled States, \textit{arXiv:1107.5245v1.} (2011).

\item \label{ref2} Yanhua Shih, Quantum Imaging, \textit{arXiv:0707.0268v1}, (2007).

\item \label{ref3} T.B. Pittman, Y.H. Shih, D.V. Strekalov, and A.V. Sergienko, \textit{Phys. Rev. A} 52, R3429 (1995).

\item \label{ref4} Y.H. Shih, \textit{IEEE J. of Selected Topics in Quantum Electronics}, 9, 1455 (2003).

\item \label{ref5} Kam Wai Clifford Chan, D. S. Simon, A. V. Sergienko, Nicholas D. Hardy, Jeffrey H. Shapiro, P. Ben Dixon, Gregory A. Howland, John C. Howell, Joseph H. Eberly, Malcolm N. O'Sullivan, Brandon Rodenburg, and Robert W. Boyd, A theoretical analysis of quantum ghost imaging through turbulence, \textit{Physical Review A} 84, 04807 (2011).

\item \label{ref6} Gregory A. Howland, P. Ben Dixon, and John C. Howell, Photon-Counting Compressive Sensing Lidar for 3D Imaging, \textit{Applied Optics,} 50, 5917 - 5920 (2011).

\item \label{ref7} P. Benjamin Dixon, Quantum Imaging and Information, \textit{Ph.D Thesis}, University of Rochester, Rochester, New York, (2011).

\item \label{ref8} C. K. Law and J. H. Eberly, Analysis and Interpretation of High Transverse Entanglement in Optical Parametric Down Conversion, \textit{Phys. Rev. Lett.} 92, 127903 (2004).

\item \label{ref9} V. Boyer, A. M. Marino, R. C. Pooser, and P. D. Lett, Entangled Images from Four-Wave Mixing, \textit{Science} 321, 544 (2008).

\item \label{ref10} S. P. Walborn, D. S. Lemelle, D. S. Tasca, and P. H. Souto Ribeiro, Schemes for quantum key distribution with higher-order alphabets using single-photon fractional Fourier optics, \textit{Phys. Rev. A} 77, 062323 (2008).

\item \label{ref11} O. Katz, Y. Bromberg, and Y. Silberberg, Compressive ghost imaging, \textit{Appl. Phys.} \textit{Lett}. 95, 131110 (2009).

\item \label{ref12} J. H. Shapiro, Computational ghost imaging, \textit{Phys. Rev. A} 78, 061802 (2008).

\item \label{ref13} I. Vidal, E. J. S. Fonseca, and J. M. Hickmann, Super-resolution quantum interference pattern of a distributed object,\textit{ Phys. Rev. A} 82, 043827 (2010).

\item \label{ref14} R. Simon, Peres-Horodecki Separability Criterion for Continuous Variable Systems, \textit{Phys. Rev. Lett. }84, 2726 (2000).

\item \label{ref15} M. V. Fedorov, M. A. Efremov, A. E. Kazakov, K. W. Chan, C. K. Law, and J. H. Eberly, Packet narrowing and quantum entanglement in photoionization and photodissociation, \textit{Phys. Rev. A} 69, 052117 (2004).

\item \label{ref16} C. K. Law and J. H. Eberly, Analysis and Interpretation of High Transverse Entanglement in Optical Parametric Down Conversion, \textit{Phys. Rev. Lett. }92, 127903 (2004).

\item \label{ref17} B. I. Erkmen and J. H. Shapiro, Unified theory of ghost imaging with Gaussian-state light, \textit{Phys. Rev. A} 77, 043809 (2008). 

\item \label{ref18} G. Scarcelli, V. Berardi, and Y. Shih, Phase-conjugate mirror via twophoton thermal light imaging, \textit{Appl. Phys. Lett}. 88, 061106 (2006). 

\item \label{ref19} F. Ferri, D. Magatti, A. Gatti, M. Bache, E. Brambilla, and L. A. Lugiato, High-Resolution Ghost Image and Ghost Diffraction Experiments with Thermal Light, \textit{Phys. Rev. Lett. }94, 183602 (2005).

\item \label{ref20} J. Cheng, Ghost imaging through turbulent atmosphere, Opt. Express 17, 7916 (2009).

\item \label{ref21} C. Li, T. Wang, J. Pu, W. Zhu, and R. Rao, Ghost imaging with partially coherent light radiation through turbulent atmosphere, \textit{Applied Physics B:Lasers and Optics} 99, 599 (2010).

\item \label{ref22} P. Zhang, W. Gong, X. Shen, and S. Han, Correlated imaging through atmospheric turbulence, \textit{Phys. Rev. A} 82, 033817 (2010).

\item \label{ref23}  R. E. Meyers, K. S. Deacon, and Y. Shih, Turbulence-free ghost imaging, \textit{Appl. Phys. Lett}. 98, 111115 (2011).

\item \label{ref241} K. W. C. Chan, D. S. Simon, A. V. Sergienko, N. D. Hardy, J. H. Shapiro, P. B. Dixon, G. A. Howland, J. C. Howell, J. H. Eberly, M. N. O'Sullivan, et al., Theoretical analysis of quantum ghost imaging through turbulence, \textit{Phys. Rev. A} 84, 043807 (2011).

\item \label{ref25} L. Chen, J. Leach, B. Jack, M. J. Padgett, S. Franke-Arnold, and W. She, High-dimensional quantum nature of ghost angular Young's diffraction, \textit{Phys. Rev. A} 82, 033822 (2010).

\item \label{ref26} G. Lima, L. Neves, R. Guzman, E. S. Gomez, W. A. T. Nogueira, A. Delgado, A. Vargas, and C. Saavedra, Experimental quantum tomography of photonic qudits via mutually unbiased basis,\textit{ Opt. Express} 19, 3542 (2011).

\item \label{ref27} S. P. Walborn, B. G. Taketani, A. Salles, F. Toscano, and R. L. de Matos Filho, Entropic Entanglement Criteria for Continuous Variables,\textit{ Phys. Rev. Lett.} 103, 160505 (2009).

\item \label{ref28} S. P. Walborn, A. Salles, R. M. Gomes, F. Toscano, and P. H. Souto Ribeiro, Revealing Hidden Einstein-Podolsky-Rosen Nonlocality, \textit{Phys. Rev. Lett.} 106, 130402 (2011).

\item \label{ref29} S. Imre and L. Gyongyosi. \textit{Advanced Quantum Communications - An Engineering Approach.} Wiley-IEEE Press (New Jersey, USA), (2012).

\item \label{ref30} R. M. Gray and J. W. Goodman, \textit{Fourier Transforms,} Kluwer, (1995)

\item \label{ref31} R. Strichartz, \textit{A Guide to Distribution Theory and Fourier Transforms}, CRC Press, (1994)

\item \label{ref32} Lecture notes in Linear systems and optimization: the Fourier transform and its applications, Stanford University, online material:\\http://see.stanford.edu/see/materials/lsoftaee261/handouts.aspx (2007)

\item \label{ref33} L. Hanzo, H. Haas, S. Imre, D. O'Brien, M. Rupp, L. Gyongyosi. Wireless Myths, Realities, and Futures: From 3G/4G to Optical and Quantum Wireless, \textit{Proceedings of the IEEE, Volume: 100, Issue: Special Centennial Issue}, pp. 1853-1888. (2012).

\item \label{ref34} L. Gyongyosi, Quantum Imaging of High-Dimensional Hilbert Spaces with Radon Transform, \textit{Frontiers in Optics 2014} (2014).

\item \label{ref35} S. Lloyd, Capacity of the noisy quantum channel, \textit{Phys. Rev. A}, vol. 55, pp. 1613--1622, (1997).
\end{enumerate}

\newpage

\appendix
\setcounter{table}{0}
\setcounter{figure}{0}
\setcounter{equation}{0}
\renewcommand{\thetable}{\Alph{section}.\arabic{table}}
\renewcommand{\thefigure}{\Alph{section}.\arabic{figure}}
\renewcommand{\theequation}{\Alph{section}.\arabic{equation}}

\setlength{\arrayrulewidth}{0.1mm}
\setlength{\tabcolsep}{5pt}
\renewcommand{\arraystretch}{1.5}
\section{Appendix}
\subsection{Notations}
The notations of the manuscript are summarized in \tref{tab1}.

\begin{longtable}{|p{1.6in}|p{2.8in}|} 
\caption{Summary of notations.}
\label{tab1}
\endfirsthead
\endhead
\hline 
\textit{Notation} & \textit{Description} \\ \hline 
$\mu \left(x,p\right)$ & An unknown internal function, \textit{x} stands for the position basis, \textit{p} is the momentum basis. \\ \hline 
$\int_{M_{\phi ,i}}{\mu \left(x,p\right)}dxdp$ & An abstracted line in the high-dimensional Hilbert space. An encoded slice $\mathcal{E}\left(\mu \left(x,p\right)\right)$ of the partial mutual information function at a $\phi $, \textit{x} and \textit{p} are the position and momentum components. \\ \hline 
\footnotesize
$ \begin{array}{l}
\int_{M_{\phi \mathrm{,1}}}{\mathrm{\dots }} \\ 
\mathrm{\dots }\int_{M_{\phi ,n}}{\mu \left(x,p\right)}dxdp \end{array}
$ & Conveys the encoded partial mutual information $\mathcal{E}\left( {{\mu }_{\phi }}\left( x,p \right) \right)$. Collection of \textit{n} slices at a fixed $\phi $. \\ \hline 
\footnotesize
$ \begin{array}{l}
\int_{\phi }{\int_{M_{\phi \mathrm{,1}}}{\mathrm{\dots }}} \\ 
\mathrm{\dots }\int_{M_{\phi ,n}}{\mu \left(x,p\right)dxdp}d\phi  \end{array}
$ & The encoded full mutual information function $\mathcal{E}\left(\int_{\phi }{{\mu }_{\phi }\left(x,p\right)}d\phi \right)$. \\ \hline 
$M_{\phi }$ & Collection of \textit{n} abstracted lines that defines the partial mutual information function $I_{M_{\phi }}\left(A\mathrm{:}B\right)$ at a given $\phi $. Represents \textit{n} coincidence measurements, evaluated as $M_{\phi }\mathrm{=}\sum_n{M_{\phi ,i}}$. \\ \hline 
$I_0\left(A\mathrm{:}B\right)$ & Mutual information in the standard setting. (i.e., $\phi \mathrm{=0}$, with no Radon transforming in the post-processing) \\ \hline 
$I_{M_{\phi }}\left(A\mathrm{:}B\right)$ & Partial mutual information, extracted at a given $\phi $, evaluated as $I_{M_{\phi }}\left(A\mathrm{:}B\right)\mathrm{=}\int_{\phi }{{\mu }_{\phi }\left(x,p\right)}dxdpd\phi $. \\ \hline 
$I_{\mathcal{R}}\left(A\mathrm{:}B\right)$ & Full mutual information function in Radon transform, taken at $\mathrm{0}\mathrm{\le }\phi \mathrm{<}\pi $. \\ \hline 
$f\left(x_A,x_B\right)$ & Biphoton wavefunction in the position basis. \\ \hline 
$\rho $ & The difference of the components from the abstracted origin ${\rho }_{\mathrm{1}}$. For a given $\mu \left(x,p\right)$ at $\phi $, it is $\rho \mathrm{=}\left(x,p\right)\mathrm{\cdot }\left(\mathrm{cos}\phi \mathrm{,sin}\phi \right)\mathrm{=}x\mathrm{cos}\phi \mathrm{+}p\mathrm{sin}\phi $.  \\ \hline 
${\rho }_i$ & The ${\rho }_i$ parameter of the \textit{i}-th abstracted line. The first parameter, ${\rho }_{\mathrm{1}}$, identifies the imaginary origin in the position-momentum space. \\ \hline 
${\rho }_{\phi }$ & The collection of \textit{n} ${\rho }_i$-s each belong to a given slice $\mu \left(x_i,p_i\right)$. \\ \hline 
$f\left(p_A,p_B\right)$ & Biphoton wavefunction in the momentum basis. \\ \hline 
$\mathrm{2}w_{\mathrm{1}}$ & Gaussian width in the $x_{\mathrm{1}}\mathrm{-}x_{\mathrm{2}}$ direction. \\ \hline 
$w_{\mathrm{2}}$ & Gaussian width in the $x_{\mathrm{1}}\mathrm{+}x_{\mathrm{2}}$ direction. \\ \hline 
${\sigma }_s$\newline  & Single photon width, ${\sigma }^{\mathrm{2}}_s\mathrm{=}w^{\mathrm{2}}_{\mathrm{2}}\mathrm{+}{\left(\frac{w_{\mathrm{1}}}{\mathrm{2}}\right)}^{\mathrm{2}}$. \\ \hline 
${\sigma }_C$\newline  & Conditional width, ${\sigma }^{\mathrm{2}}_C\mathrm{=}\frac{\mathrm{4}w^{\mathrm{2}}_{\mathrm{1}}w^{\mathrm{2}}_{\mathrm{2}}}{\mathrm{4}w^{\mathrm{2}}_{\mathrm{2}}\mathrm{+}w^{\mathrm{2}}_{\mathrm{1}}}$. \\ \hline 
$d$ & Measurement dimension. Stands for the measurement space of position and momentum bases. (Practically, it represents the resolution of the measurement device in pixels.) \\ \hline 
$\mathcal{R}\left({\mu }_{\phi }\left({\rho }_{\phi },\phi \right)\right)$ & Radon transform of the function ${\mu }_{\phi }\left({\rho }_{\phi },\phi \right)$. \\ \hline 
$F^{\mathrm{-}\mathrm{1}}$ & Inverse Fourier transform. \\ \hline 
$\phi \mathrm{,0}\mathrm{\le }\phi \mathrm{<}\pi $ & Phase rotation, used by the PM (Phase Modulator.) \\ \hline 
$F_{\rho }\left(\mathcal{R}\left({\mu }_{\phi }\right)\right)$ & Fourier transform of $\mathcal{R}\left({\mu }_{\phi }\right)$ with respect to $\rho $. \\ \hline 
$\mathrm{Pr}\left(M^A_{\phi },M^B_{\phi }\right)$ & Joint detection probability of measurements $M^A_{\phi }\mathrm{=}\sum_a{M^A_{\phi ,i}}$ and $M^B_{\phi }\mathrm{=}\sum_b{M^B_{\phi ,i}}$, at a given $\phi $, with respect to the position basis, as\newline $\mathrm{Pr}\left(M^A_{\phi },M^B_{\phi }\right) \newline \mathrm{=}\int_{M^A_{\phi }}{dx_A}\int_{M^B_{\phi }}{dx_B}{\left|f\left(x_A,x_B\right)\right|}^{\mathrm{2}}$. \\ \hline 
$\mathrm{Pr}\left(M^A_{\phi },M^B_{\phi }\right)$ & Joint detection probability of measurements $M^A_{\phi }\mathrm{=}\sum_a{M^A_{\phi ,i}}$ and $M^B_{\phi }\mathrm{=}\sum_b{M^B_{\phi ,i}}$, at a given $\phi $, with respect to the momentum basis, as\newline $\mathrm{Pr}\left(M^A_{\phi },M^B_{\phi }\right) \newline \mathrm{=}\int_{M^A_{\phi }}{dp_A}\int_{M^B_{\phi }}{dp_B}{\left|f\left(p_A,p_B\right)\right|}^{\mathrm{2}}$. \\ \hline 
$\mathrm{Pr}\left(M^A_{\phi },\phi \right)$ & Detection probability for measurement $M^A_{\phi }\mathrm{=}\sum_a{M^A_{\phi ,i}}$, at a given $\phi $, as\newline $\mathrm{Pr}\left(M^A_{\phi },\phi \right)\mathrm{=}\sum_{M^B_{\phi }}{\mathrm{Pr}\left(M^A_{\phi },M^B_{\phi }\right)}$. \\ \hline 
$\mathrm{Pr}\left(M^B_{\phi }\right)$ & Detection probability for measurement $M^B_{\phi }\mathrm{=}\sum_b{M^B_{\phi ,i}}$, at a given $\phi $, as\newline $\mathrm{Pr}\left(M^B_{\phi }\right)\mathrm{=}\sum_{M^A_{\phi }}{\mathrm{Pr}\left(M^A_{\phi },M^B_{\phi }\right)}$. \\ \hline 
$\boldsymbol{\mathrm{SB}}$ & Entropic separability bound. \\ \hline 
$\boldsymbol{\mathrm{S}}{\boldsymbol{\mathrm{B}}}_0$ & Entropic separability bound in the standard model. \\ \hline 
$\boldsymbol{\mathrm{S}}{\boldsymbol{\mathrm{B}}}_{\mathcal{R}}$ & Entropic separability bound under Radon transform. \\ \hline 
$\left(r,\phi \right)$ \textit{} & Polar coordinates in the $\left({\lambda }_{\mathrm{1}},{\lambda }_{\mathrm{2}}\right)$ plane, where ${\lambda }_{\mathrm{1}}\mathrm{=}r\mathrm{cos}\phi $, ${\lambda }_{\mathrm{2}}\mathrm{=}r\mathrm{sin}\phi $, and \textit{r} is the imaginary frequency parameter, $r^{\mathrm{2}}\mathrm{=}{\lambda }^{\mathrm{2}}_{\mathrm{1}}\mathrm{+}{\lambda }^{\mathrm{2}}_{\mathrm{2}}$. \\ \hline 
$\mathcal{C}\left(\mathrm{\cdot }\right)$\textit{} & Cartesian representation of a polar grid point $\chi $. \\ \hline 
$\mathcal{C}\left(\chi \right)$ & Cartesian data points calculated by the weighted average of the $\chi $ polar grid points as \newline $\mathcal{C}\left(\chi \right)\mathrm{=}w_{\mathrm{1}}\left({\chi }_{\mathrm{1}}\right)\mathrm{+}w_{\mathrm{2}}\left({\chi }_{\mathrm{2}}\right)\mathrm{+}w_{\mathrm{3}}\left({\chi }_{\mathrm{3}}\right)\mathrm{+}w_{\mathrm{4}}\left({\chi }_{\mathrm{4}}\right)$, where $\chi \left(r_f,{\phi }_j\right)$ is a data point in the grid. \\ \hline
\end{longtable}
\newpage
\subsection{Abbreviations}
The abbreviations of the manuscript are summarized as follows.\\
\begin{description}
\item[BS] Beam Splitter
\item[CV] Continuous-Variable
\item[CVQKD] Continuous-Variable Quantum Key Distribution
\item[PM] Phase Modulator
\item[SB] Separability Bound
\item[SPDC] Spontaneous Parametric Down-Conversion
\end{description}
\end{document}